\documentstyle[emulateapj]{article}

\def\lae{\mathrel{<\kern-1.0em\lower0.9ex\hbox{$\sim$}}}
\def\gae{\mathrel{>\kern-1.0em\lower0.9ex\hbox{$\sim$}}}

\def\tt{\tablenotetext}

\slugcomment{\smallskip Accepted for publication in the Astrophysical Journal}
 
\lefthead{Shetrone et al.}
\righthead{Abundances of Dwarf Spheroidal Galaxies}
 
\begin{document}
 
\title{Abundance Patterns in the Draco, Sextans and Ursa Minor Dwarf Spheroidal Galaxies}
 
\author{Matthew D. Shetrone}
\affil{McDonald Observatory, University of Texas, P.O. Box 1337, Fort Davis, Texas 79734 \\
   {\rm shetrone@astro.as.utexas.edu}}

\medskip

\author{Patrick C\^ot\'e\altaffilmark{1}}
\affil{California Institute of Technology, Mail Stop 105-24, Pasadena, CA 91125, and \\
Department of Physics \& Astronomy, Rutgers University, 136 Frelinghuysen 
Road, Piscataway, NJ 08854-8019 \\
   {\rm pcote@physics.rutgers.edu}}

\and

\author{W.L.W Sargent}
\affil{California Institute of Technology, Mail Stop 105-24, Pasadena, CA 91125 \\
   {\rm wws@astro.caltech.edu}}

\medskip

\altaffiltext{1}{Sherman M. Fairchild Fellow}

 
 
\begin{abstract}
The Keck I telescope and the High Resolution Echelle Spectrometer (HIRES) have
been used to obtain spectra for red giant stars belonging to 
the Draco, Sextans and Ursa Minor dwarf spheroidal (dSph) galaxies. 
An analysis of these spectra is presented, along with abundance ratios 
for more than 20 elements. The resulting database of element 
abundances for 17 dSph stars is the most extensive yet assembled
for stars in such environments. Our principal findings are summarized as follows:
(1) There is unambiguous evidence for a large internal dispersion
in metallicity in all three galaxies: our program stars span a
range of ${\Delta}$[Fe/H] = 1.53, 1.40 and 0.73 dex in Draco, Sextans and 
Ursa Minor, respectively.
(2) The abundance patterns among the dSph stars are remarkably uniform, 
suggesting that three galaxies have similar nucleosynthetic histories
and, presumably, similar initial mass functions. 
(3) A comparison of the measured element abundance ratios for our sample 
of dSph stars with published values for Galactic halo and disk field stars
suggests that the dSph galaxies have $0.02 \lae$ [$\alpha$/Fe] $\lae$ 0.13 dex,
whereas the halo field star sample has [$\alpha$/Fe] $\sim$ 0.28 dex
over the same range in metallicity.
(4) The most metal-rich dSph stars in our sample have [Y/Fe] abundances 
which are significantly lower than those measured for halo field stars of 
similar metallicity, while the measured [Ba/Eu] ratios for the dSph stars suggest 
that the early chemical evolution of these galaxies was dominated by the r-process.
Taken together, these results suggest that the Galactic halo is unlikely
to have assembled, {\it in its entirety}, through the disruption of dwarf
galaxies similar to the low-luminosity, $\langle L_V \rangle = 3\times10^5L_{V,{\odot}}$,
dSphs studied here.
We also note that the measured [Zn/Fe] abundance ratios for the dSph stars 
exceed those of damped Ly$\alpha$ systems having low levels of dust depletion by 
roughly an order of magnitude.

The first high-resolution abundance analysis for the distant Galactic globular cluster 
NGC 2419 is also presented. From a HIRES spectrum of a single red giant, we find a metallicity of 
[Fe/H] = $-2.32\pm0.11$ dex. This is slightly lower than, but still consistent with, 
published estimates based on low-resolution spectroscopy. With the possible 
exception of a slight enhancement in the abundances of some heavy elements such
as Ce, Nd, Y and Ba, the observed abundance pattern closely resembles those exhibited
by red giants in M92: a nearby, well-studied globular cluster of nearly 
identical metallicity.
\end{abstract}
 
 
 
 
%

\section{Introduction}

As isolated, low-mass systems, dwarf galaxies are probably the closest 
approximations in Nature to idealized ``closed'' or ``leaky box'' 
models of chemical enrichment.
The old, nearby dwarf satellites of the Milky Way thus offer a unique opportunity
to study the formation and chemical evolution of galaxies in a level of detail 
which will never be possible with high-redshift systems.

The apparent dearth of gas and young stars in the Galactic dwarf spheroidal 
(dSph) galaxies, coupled with their low metallicities, led 
early researchers to regard them as similar to globular clusters in terms
of their stellar populations ($e.g.$, Hodge 1971), despite clear differences 
in their respective structural parameters. However, careful scrutiny has now revealed
these systems to be far from simple. Mateo (1998) 
recently reviewed the now overwhelming observational evidence 
for complex and varied star formation histories in these faint systems.
High-precision photometric studies, and low-resolution spectroscopy of
individual red giant branch (RGB) stars, has firmly established that both
Galactic and M31 dSph galaxies show evidence for large internal metallicity 
variations ($e.g.$, Zinn 1978; Stetson 1984; Suntzeff et al. 1993; 
C\^ot\'e, Oke \& Cohen 1999; Da Costa et al. 2000). Recently, Shetrone, Bolte \& Stetson 
(1998; hereafter SBS) presented an abundance analysis for four RGB 
stars belonging to the Draco dSph galaxy: the first study of dSph stars to 
make use of high-resolution ($R$ = 34,000) spectroscopy.
SBS measured abundance ratios for a variety of elements and, 
despite the limited sample size, found unmistakable evidence 
for a wide range in metallicity of $-3 \lae$ [Fe/H] $\lae -1.5$ dex.

Elemental abundances represent a potentially powerful means of testing the 
suggestion that dwarf galaxies are the surviving ``building blocks'' from which 
larger galaxies formed (Larson 1988; Zinn 1993).  Such tests are particularly 
topical in view of the emerging empirical and analytical
evidence that galactic halos are assembled 
from chemically-distinct, low-mass fragments ($e.g.$, Searle \& Zinn
1978; Zinn 1993; Ibata et al. 1994; Klypin et al. 1999; Moore et al. 1999; 
C\^ot\'e et al. 2000; Yanny et al. 2000). Some hydrodynamical simulations of galaxy formation 
also point to the assembly of large galaxies from low-mass, gas-rich, 
proto-galactic fragments ($i.e.$,
Haehnelt, Steinmetz \& Rauch 1998; 2000). If these gaseous fragments are the
high-redshift analogs of local dwarf galaxies, then a correspondence
between the abundance patterns of the dSphs and those of damped Ly$\alpha$
(DLA) absorbers might be expected; such a correspondence in the {\sl mean} metallicities
has been known for some time (see Pettini et al. 1997 and references therein),
although the association of DLA systems with spiral disks (Wolfe et al. 1995; Prochaska \&
Wolfe 1997) or dwarf galaxies (York et al. 1986; Tyson 1988; Pettini, 
Boksenberg \& Hunstead 1990) remains an open question. By examining the abundance
patterns of dSph galaxies, and comparing with those of DLA systems and Galactic halo/disk
field stars, it may be possible to discriminate between these
different scenarios.

In this paper, we build upon the initial study of SBS by analyzing
high-resolution spectra for an expanded sample of RGB stars belonging
to the Draco, Sextans and Ursa Minor dSph galaxies, as well as for
a single red giant in the distant globular cluster NGC 2419. We compare the 
element abundance ratios measured for our sample of dSph stars to those
of Galactic halo/disk field stars and DLA absorbers having low
levels of dust depletion.

\section{Observations and Reductions}

\subsection{Selection of Program Stars}

RGB stars belonging to three Galactic dSph galaxies ($i.e.$, Draco, Sextans 
and Ursa Minor) were selected 
from a combination of published color-magnitude diagrams, new unpublished
magnitudes (Stetson 2000, private communication and 
Cudworth 2000, private communication) and radial velocity 
catalogs ($e.g.$, Da Costa et al. 1991; Suntzeff et al. 1993; 
Stetson 1979, 1984; Armandroff, Olszewski \& Pryor 1995). Published 
radial velocities were used to ensure that only {\it bona fide} members of their 
respective galaxies were targeted for observation with HIRES,
the High Resolution Echelle Spectrometer on the Keck I 10m telescope
(Vogt et al. 1994).

Color magnitude diagrams for the three dSph galaxies studied here are shown in 
Figure 1. RGB stars for which we have HIRES spectra are indicated by the large symbols. 
These three galaxies contain little or no gas (Blitz \& Robishaw 2000) and are 
comprised predominantly of old stars, although Mateo, Fischer \& Krzeminski (1995
report the presence of a modest intermediate-age population in Sextans.
The four Draco red giants observed by SBS are also shown in Figure 1, 
as we have re-analyzed their spectra here (see \S 4).  The lower right panel of Figure 1 
shows the distribution of our sample of RGB stars in the $M_V$-$(B$-$V)_0$ plane. 
The dashed lines in this panel show 13 Gyr isochrones from Bergbusch \& VandenBerg 
(1992) having metallicities of [Fe/H] = $-2.26$, $-1.66$ and $-1.26$ dex.

\subsection{HIRES Spectroscopy}

HIRES spectra for 13 RGB stars belonging to the Draco, Ursa Minor and Sextans dSph galaxies 
were acquired during three observing runs at the Keck I telescope. During the March
1999 observing run, we also obtained a spectrum of a single red giant in the outer-halo
globular cluster NGC 2419. A number of bright RGB stars belonging to the nearby, well-studied
globular clusters M92 and M3 were also observed during the March 1999 and July 1999
observing runs, in order to compare our measured abundances with those of previous
high-resolution spectroscopic studies. An observing log for these observations is 
presented in Table 1. For each run, we used the C5 decker to produce a 
1\farcs15$\times$7$^{\prime\prime}$ slit and a spectral resolution of 
${\lambda} / {\Delta}{\lambda}$ = 34,000. In all cases, the detector was binned 1$\times$2
in the spatial direction in order to reduce the read noise. 
Spectra obtained during the July 1999 and August 1999 observing runs span the
wavelength region $4540 \lae \lambda\lambda \lae 7020$ \AA ; different echelle and cross 
disperser angles were used during the March 1999 observations, giving a wavelength coverage
of $3850 \lae \lambda\lambda \lae 6300$ \AA .
The raw spectra were reduced 
using the MAKEE software package (Tom Barlow, private communication), and wavelength-calibrated 
within the IRAF\altaffilmark{2}\altaffiltext{2}{IRAF is distributed by the National Optical Astronomy 
Observatories, which are operated by the Association of Universities for Research in Astronomy, 
Inc., under contract to the National Science Foundation.} environment.

Table 2 summarizes several properties of the program stars.  From left to right, this table 
records the star name, $V$ magnitude, $(B-V)$ color, interstellar extinction, HIRES exposure 
time, signal-to-noise ratio (S/N) measured at the continuum near $\lambda \simeq$ 6100 $\rm \AA$, 
and heliocentric Julian date. The final column
records the heliocentric radial velocity of each star, measured directly from the 
spectra with the velocity zero-point established using telluric lines (see Shetrone 1994).
References for the photometry are given in the notes to the table. For all objects,
the adopted reddening is taken from the DIRBE reddening maps of 
Schlegel, Finkbeiner \& Davis (1998).

\section{Analysis}

The lines chosen for the abundance analysis were adopted from several 
sources, including Blackwell et al. (1982), Blackwell et al. (1986), 
Bizzarri et al. (1993), Fuhrmann et al. (1995),
McWilliam et al. (1995; M95), Kraft et al. (1995), Shetrone
(1996), Sneden et al. (1996), Carretta \& Gratton (1997), 
and the National Institute of Standards
and Technology Atomic Spectra Database.  In some cases, the line choices 
differ from those of SBS, mainly in the addition of more and
better lines for Ti I, Ti II, Mg I and several rare earth elements.
Equivalent widths (EWs) were measured 
from Gaussian fits to individual spectra lines. Table 3 lists the measured 
EWs and adopted line parameters for each of the elements considered below.

Figure 2 shows a comparison of our EWs for the M92 and M3 RGB stars which we 
have in common with Sneden et al. (1991) and Kraft et al. (1992).
There is a essentially no offset, $\Delta$EW = $1.6{\pm}1.2$ m\AA , 
between the two datasets, and the rms scatter of ${\sigma}$ = $7.1$ m$\rm \AA$ 
is consistent with the errors expected from uncertainties of the individual 
measurements. These results are consistent with those found in SBS.    

Initial temperature estimates were made using the colors and reddenings from 
Table 2, using the de-reddened colors, $(B-V)_0$, and our own 
$(B-V)_0$-T$_{\rm eff}$ calibration for red giants with metallicities in
the range $-3.0 \le$ [Fe/H] $\le -1.0$ dex. These initial estimates 
were based on rough approximations of the metallicities of the program stars 
based upon their location in the color magnitude diagrams.  This estimate was 
then fine-tuned in two ways. First, the metallicity from the
first iteration was fed back into our color temperature estimate 
and the entire process was repeated until we converged upon a best color 
temperature. Second, we adjusted the color temperature by: (1) demanding a minimized 
slope, to within the errors, in the plots of Fe abundance (from Fe I) versus both
excitation potential and equivalent width and; (2) requiring that the abundance of
the {\it ionized} species equal that of the {\it neutral} species (based largely
upon Fe I and Fe II, and to a lesser extent, Ti I and Ti II).   
In making this minimization, the microturbulent velocity, 
effective temperature and surface gravity were adjusted iteratively.  

Among the sample of reference stars in M92 and M3, the largest deviation 
between the initial and final adopted parameters were
${\Delta}{\rm T}_{\rm eff}$ = 75 K and ${\Delta}{\log{g}}$ = 0.10 dex for the 
effective temperature and surface gravity, respectively.  Even in the case of the much 
lower S/N spectrum of the red giant RH-10 in NGC 2419, the deviations were only 
$-25$ K and 0.1 dex.  For the sample of dSph stars, the largest deviation
from the predicted temperature was 175 K for Ursa Minor K.   This 
giant has obvious C$_2$ bands which reveal it to be 
a carbon star; it is not surprising that the color temperature
based upon non-carbon stars would predict a significantly cooler temperature
than the actual spectroscopic effective temperature.   Excluding this
star, the average differences between the predicted and adopted temperatures 
were +63$\pm$18, +38$\pm$25 and
+22$\pm$18 K in Ursa Minor, Draco and Sextans, respectively.  The 
differences in the predicted versus adopted surface gravities are 
$-0.1\pm0.1$ $0.0\pm0.2$ and and $-0.15\pm0.1$ dex.  These slight differences
in the stellar parameters could arise in several ways:  a slight
underestimation (0.02 mag) of the true reddenings, the use of a halo star 
$(B-V)_0$-T$_{\rm eff}$ relation for the dSph giants (see \S5.1), systematic
errors in the unpublished photometry, or
a systematic error due to working with low-S/N spectra.
We reject this last possibility because the predicted and adopted 
parameters for NGC 2419 RH-10, which has a lower S/N ratio
than the majority of the dSph spectra, are in excellent agreement,
while the dSph giants having the highest S/N spectra ($i.e.$, those in Ursa 
Minor) exhibit the largest deviations.

Model atmospheres were taken from the computations of the MARCS code
(Gustafsson et al. 1975), and the abundance calculations were performed
using the current version of Sneden's (1973) LTE line analysis 
and spectrum synthesis code.  The procedures are nearly identical to
those employed in SBS, except for the modified line list.  
Since some important lines were added, we have re-analyzed
the Draco giants from SBS and included the results in this paper.  It is important
to note that, as with most high-resolution spectroscopic
analyses of metal-poor stars using the MARCS models, we have compensated
for the increased electron contribution from the overabundant alpha elements 
by slightly increasing the overall metallicity (+0.1 dex) of the model 
atmospheres. This compensation for alpha element enhancement will be 
discussed further in \S5.1.  The best-fit model parameters for each 
star, along with the measured EWs, are recorded in Table 3.

Note that several of the elements in our analysis ($e.g.$, V, Mn, Ba, Eu) are known 
to exhibit hyper-fine splitting.   We have not included this hyper-fine 
splitting in our analysis because our goal is the measurement of
overall abundance {\it differences}, not absolute abundances. The 
magnitude of the correction is small in comparison to the EW measurement 
errors: $i.e.$, for [Ba/Fe] the effect is only 0.02 dex. 

\section{Abundances}

The first important result of this work is the overall abundance spread within
the three dSph galaxies.  In Figure 1, the alpha-enhanced isochrones of Bergbusch \&
VandenBerg (1992) are overlaid on the color-magnitude diagrams, with our sample 
of dSph stars indicated.  If the galaxies are comprised primarily of old and coeval 
stars, and if the photometric uncertainties are negligible, then a metallicity range of 
$-2.5 \lae$ [Fe/H] $\lae -1.5$ dex can be expected in all three galaxies. As Table 4 
shows, our measured abundances are roughly consistent with this expectation:
$i.e.$, we find metallicity ranges of $\Delta$[Fe/H] $\simeq$ 1.53, 1.40 and
0.73 dex, respectively, for Draco, Sextans and Ursa Minor. Moreover, the stars
with the lowest abundances are found close to the blue edge of the color
distribution, as expected. A more detailed isochrone-age analysis is beyond the
scope of this paper since, as discussed in \S5.1, the current set of
alpha-enhanced isochrones may not be entirely appropriate for these galaxies.

The weighted mean metallicities of the three dSph
samples are $\langle$[Fe/H]$\rangle$ = $-2.00\pm0.21$, $-1.90\pm0.11$ and $-2.07\pm0.21$ dex 
for Draco, Ursa Minor and Sextans, respectively. Despite the small sample sizes,
and the presence of large internal metallicity spreads
within all three galaxies, these mean values are in good agreement with previous
determinations (see, $e.g.$, Mateo 1998).
The mean metallicity of the lone red giant in NGC 2419 is found to be 
[Fe/H] = $-2.32\pm0.11$ dex. This is
slightly lower than, but still consistent with, the value of
[Fe/H] = $-2.10\pm0.15$ dex found by Suntzeff, Kraft \& Kinman (1988) from
low-resolution spectroscopy of six red giants.

\subsection{Light Elements}

Many of the light elements such as O, Na, Mg, Al are known to exhibit significant 
star-to-star variations in globular clusters. For instance, 
Shetrone (1996), Kraft et al. (1997) and Sneden et al. (1997) find
large ranges in the abundances of these light elements for red giants belonging to 
M3 and M92: $-0.7$ $<$ [O/Fe] $<$ +0.45 dex; $-0.3$ $<$ [Na/Fe] $<$ +0.5 dex;
$-0.1$ $<$ [Mg/Fe] $<$ +0.4 dex; and 0.0 $>$ [Al/Fe] $>$ +1.1.
By contrast, halo field stars do not exhibit these abundance variations.
The observed variations in the globular cluster stars follow a specific pattern 
which is sometimes referred to as a ``deep mixing abundance pattern'', with
Al and Na being enhanced, and O (and sometimes Mg) being depleted.
Any comparison of the dSph abundance patterns for these elements with those 
of the globular cluster stars is complicated by these variations. 
In Figure 3, the light element abundances for our program stars are shown 
along with a sample of Milky Way halo and disk stars.  At least
three of the M92 giants and one of the M3 giants exhibit the aforementioned
``deep mixing abundance pattern''.  Unfortunately, most of the O, Na and
Al abundances for the dSph stars are upper limits, making a detailed
comparison with the field star sample impossible.

\subsection{Even-Z ($\alpha$) Elements}

The light elements with even numbers of protons ($e.g.$, O, Mg, Si, Ca, Ti)
are sometimes referred to as $\alpha$ elements, or even-Z elements.
Figure 4 shows our measured abundances for these even-Z elements,
except for O and Mg which are shown in Figure 3, plotted against the 
iron abundance. The abundances of the globular cluster stars ---
M3, M92 and NGC 2419 --- are consistent with those of the halo field
stars.  The Galactic halo field stars exhibit $\alpha$-enhancements 
between 0.1 and 0.5 dex (with the exception
of [Si/Fe] which shows a much broader distribution)
over the metallicity range $-3.0 \lae$ [Fe/H] $\lae -1.2$ dex.  Due to the large
intrinsic spread among the Si abundances, and because of our rather large
errors and upper limits on the measured Si abundances, the [Si/Fe] abundance
pattern will be discussed no further. The bottom panel in
Figure 4 gives {\it average} even-Z abundances, 
[${\alpha}$/Fe] = ${1 \over 3}$([Mg/Fe + Ca/Fe + Ti/Fe]),
for both our program stars and for those stars taken from the literature. This
definition excludes the contribution from [O/Fe] since many of the stars
(both in our sample and in the literature) do not have detected O lines.

A first impression from Figure 4 is that the stars belonging to the three 
different dSph galaxies occupy the same portion of the figures: $i.e.$, they 
have roughly the same abundance patterns.  In addition, their abundance
pattern appears to differ from that of the Galactic halo field star sample,
with the dSph stars falling below the Milky Way sample at a given 
metallicity. Considering the dSph samples separately, the average 
even-Z abundances, [$\alpha$/Fe], are: 0.09$\pm$0.02 dex for 
Draco, 0.13$\pm$0.04 dex for Ursa Minor, and 0.02$\pm$0.07 dex 
for Sextans.  Over the same range in metallicity,
the halo field star sample has a mean value of [$\alpha$/Fe] = 0.28$\pm$0.02 dex.
Thus, all three dSph samples have a statistically 
significant underabundance of Mg, Ca and Ti in comparison to the 
Galactic halo. By contrast, we find [$\alpha$/Fe] $\simeq$ 0.21$\pm$0.10 dex
for the lone red giant in NGC 2419, which is slightly lower than, but 
nevertheless consistent with, the mean value of 0.34$\pm$0.04 dex
for the red giants in M92.

There is also some evidence for a trend of decreasing [$\alpha$/Fe] abundance with 
increasing metallicity in the Sextans and Ursa Minor samples.  The Ursa Minor sample 
is linearly correlated with a slope of d[$\alpha$/Fe]/d[Fe/H] = $-0.26\pm0.11$. The
trend is less obvious among sample of Sextans stars, which have a slope of $-0.12\pm0.08$. 
Nonetheless, it is certainly true that the most metal-rich star in the Sextans
sample has a significantly lower even-Z abundance than the rest
of the Sextans sample. Similarly, the most metal-rich star in the Draco 
sample has the lowest [O/Fe] and [Mg/Fe] abundance ratios for this galaxy.

\subsection{Iron Peak Elements}

Figures 5 and 6 show the iron peak element abundance ratios plotted 
against metallicity.  The [V/Fe] abundances are under-sampled due to 
the temperature sensitive nature of these lines: $i.e.$, the 
more metal-poor stars are hotter and hence the V lines are weaker 
due to both the lower abundances and the cooler temperatures.
All of the dSph giants with measured V abundances exhibit the same 
abundance pattern as that of the halo field star sample, with the exception of 
Draco 473.  This may be a further indication of chemical 
peculiarity of this metal-rich star, or it could indicate a problem 
with our adopted temperature.

Among the halo field stars more metal rich than ${\rm [Fe/H]}~\simeq -1.9$ dex,
Co and Cr are found in their solar ratios; at lower metallicities,
the abundances of Co and Cr diverge, with the [Co/Fe] ratios increasing
and the [Cr/Fe] ratios decreasing. This same abundance pattern is seen 
in each of the three dSph samples. Like [Cr/Fe], the [Cu/Fe] and 
[Mn/Fe] abundances exhibit a decline with decreasing iron abundance, although
the onset occurs at higher metallicity. All three dSph samples 
exhibit sub-solar [Cu/Fe] and [Mn/Fe] abundance ratios, consistent with 
that found in the Milky Way sample. In addition, both [Ni/Fe] and [Zn/Fe] 
are found in their solar ratios in the halo and dSph samples over the 
metallicity range of interest.

For NGC 2419, the measured [Cr/Fe], [Mn/Fe], [Co/Fe], [Ni/Fe] and [Cu/Fe]
abundances are similar to those found for the red giants in M92.

\subsection{Heavy Metals}

We define the heavy metals as those elements with ${\rm Z} > 30$.   This broad
definition includes many subcategories including the first s-process elements,
the second s-process elements, and the r-process elements.   The s-process
elements are those produced mainly by slow neutron addition, while the 
r-process elements are created largely through the rapid addition of neutrons.
Table 4 of Burris et al. (2000) gives the relative contributions of the s-
and r-processes for for the all of the heavy elements in the sun.
Of the heavy metals for which we have measured abundances, we have one 
first s-process peak element (Y), three second s-process peak elements 
(Ba, Ce, Sm), one r-process element (Eu) and one element which is nearly 
an even mix of the s- and r-processes (Nd). Figures 7 and 8 show the heavy 
metal abundances for the dSph, halo and globular cluster samples.

While the concept of s-process elements and r-process elements is a 
traditional one, it can lead to possible misconceptions.   In most 
Population II stars, the r-process dominates and the s-process 
contributes little to the abundance of the ``second s-process peak elements''.
A notable exception to this trend is found among carbon stars.   These asymptotic 
giant branch (AGB) stars have contaminated atmospheres, with anomalously high
fractions of carbon and s-process elements on their surfaces. As a result,
their spectra are rich in C$_2$ bands and the lines of s-process elements.
One star in our sample, Ursa Minor K, has a spectrum with strong C$_2$ bands,
as noted previously from lower resolution spectroscopy (Canterna \& Schommer
1978, Aaronson et al. 1983, Armandroff,
Olszewski \& Pryor 1995). Our analysis reveals an abundance pattern for
this star which is dominated by the s-process, and confirms the 
classification of Ursa Minor K as a carbon star.

The abundance ratio [Ba/Eu] is most often used to access the relative 
contribution of the r- and s-processes to the heavy metal abundance pattern.   
Figure 8 shows the measured [Ba/Eu] ratios plotted against 
metallicity for our program stars.  For the halo field star sample, the [Ba/Eu] abundances range 
from solar-like ratios in the most metal-rich stars, to [Ba/Eu] $\simeq -0.5$ 
for [Fe/H] $\lae -1$ dex.   This trend can be understood as a evolution from 
old metal-poor stars having abundance patterns dominated by the r-process, to
younger solar-metallicity stars with a mix of r-process and s-process patterns.
Unfortunately, the spectrograph setup used during the observation of stars in Sextans 
did not include the lone Eu line which was included in the Draco and Ursa 
Minor spectra. For this reason, we do not have a [Ba/Eu] ratio for the Sextans
stars, and hence no information about the relative r- and s-process contributions.
However, the Draco and Ursa Minor stars (Ursa Minor K excluded) exhibit 
the same r-process dominated abundance pattern as does the Milky Way sample.

Inspection of Table 4 and Figures 7 and 8 reveals three stars whose heavy 
element abundance ratios are enhanced relative to those typical for other
dSph and halo field stars: Ursa Minor K, Ursa Minor 199 and Sextans 35.  As mentioned above, 
Ursa Minor K is an obvious carbon star with an enhanced s-process dominated 
abundance pattern, while the the heavy element abundance pattern for Ursa Minor 199 is
dominated by the r-process. Unfortunately, we do not know what process dominates 
the heavy element abundance pattern for Sextans 35, since we lack a measured Eu abundance
for this star.
The enhancement of the heavy element abundances for these stars does have
a precedent among halo field stars; a small percentage of Population II 
stars which show r-process dominated heavy element abundance patterns have 
large over abundances of the heavy elements with respect to iron ($e.g.$, 
Westin et al. 2000, Norris et al. 1997, Cowan et al. 1995)

Excluding the three dSph stars with large overabundances of the heavy
elements, the abundances of the second s-process elements (Ce, Sm, and Ba)
are consistent with those of the halo field stars. This is also true 
for the abundances of the r-process peak element Eu. However, the 
abundances of Y (a first s-process peak element) are much lower in the more
metal-rich dSph giants than in the halo field stars of similar metallicity.
In the bottom panel of Figure 8, we have plotted the [Ba/Y] abundance ratio
against the metallicity.  It is clear that the dSph sample has a  significantly 
higher [Ba/Y] abundance ratio than the halo field star sample over the entire 
range of dSph metallicities.

For NGC 2419, the measured ratios of [Ce/Fe], [Nd/Fe], [Y/Fe] and [Ba/Fe]
appear to be slightly enhanced over their respective ratios in the M92
giants. Interestingly, the inferred ratio of [Ba/Y] = $-0.13\pm0.14$ is 
in good agreement with the corresponding values measured for red giants in M92.

\section{Discussion}

The abundance patterns of the three dSphs sampled here are remarkably uniform. 
This suggests that the galaxies share fairly similar nucleosynthetic histories, 
and thus, similar initial mass functions.  Because dwarf galaxies (and DLA
systems) have been invoked in numerous models for the formation of large 
galaxies, we now address the issue of how Sextans, Draco and Ursa Minor as a 
group fit into these scenarios.

\subsection{Milky Way Comparison}

As mentioned above, there are some significant differences
between the abundance patterns in the Milky Way and the dSph samples.
Probably the most important of these is the lower even-Z abundances found 
among the dSph stars. Since the production of even-Z elements is thought to 
be dominated by massive Type II supernovae ($e.g.$, Tsujimoto, et al. 1995),
lower [$\alpha$/Fe] abundance ratios would require either that the most massive 
Type II supernovae were absent in the young dSphs, or that the ejecta from these
massive supernovae were lost from the galaxy and not 
incorporated in the subsequent generations of stars.  Alternatively, it
is possible that the chemical evolution of the dSph galaxies included a
relatively large contribution from Type Ia supernovae (which are
expected to produce large quantities of iron peak elements compared to 
the $\alpha$-elements).

The importance of the low [$\alpha$/Fe] abundance ratios should not be
understated.  The even-Z elements are abundant electron donors and hence important
sources of atmospheric opacity in K giants; thus, the abundances of even-Z elements
influence age estimates based upon isochrone fitting ($e.g.$, Bergbusch \& 
VandenBerg 1992). As mentioned in \S3, 
the analysis of halo stars includes the extra electron contribution to the 
opacities due to the even-Z elements, which is offset by artificially 
inflating the metallicity of the model atmosphere. We have followed this 
same procedure for the dSph stars; if this compensation were not included, 
the abundances of the neutral species would {\it increase} by 0.03 dex, 
while the abundance of the ionized species would {\it decrease}
by 0.03 dex. Since our metallicities are based mainly upon the large number of
Fe I lines, we may have underestimated the dSph abundances systematically
by 0.03 dex.  The abundance ratios for the neutral species should not 
change, but the abundance ratios reported in Table 4 for the ionized species
may be systematically too large by 0.06 dex for stars which do not have
enhanced abundances of even-Z elements.

Although the Galactic halo does contain stars with low even-Z abundance patterns
($e.g.$, Ivans et al. 2000), such stars are rare. The observed differences in the 
respective even-Z element abundance patterns may therefore
put some interesting limitations on the suggestion that the Milky way was assembled from 
``building blocks'', or proto-Galactic fragments, similar to the dSphs sampled here 
($e.g.$, Searle \& Zinn 1978; Larson 1988; Zinn 1993; Mateo 1996; C\^ot\'e et al. 2000). 
If these suggestions are correct, then perhaps the actual proto-Galactic fragments were 
larger and could better retain gas 
from massive Type II supernovae than the present sample of low-luminosity dSph galaxies. In the
Monte-Carlo simulations of C\^ot\'e et al. (2000), the initial population of 
proto-Galactic fragments span a wide range in luminosities, including a small 
number of large systems with present-day luminosities of 
$L_V \sim$ $2\times10^{8}$ $L_{V,{\odot}}$ as well as numerous small objects with
luminosities similar to those of Draco, Sextans and Ursa Minor: 
$\langle L_V  \rangle \sim 3\times10^{5}$ $L_{V,{\odot}}$.
If the chemical enrichment of the proto-Galactic fragments can be roughly approximated
by a Closed Box model ($e.g.$, Talbot \& Arnett 1971; Searle \& Sargent 1972), then 
those proto-Galactic fragments which are similar in luminosity to the Fornax dSph 
($L_V = 1.5\times10^7$ $L_{V,{\odot}}$) galaxy are expected to have
contributed roughly half of the number of halo field stars with [Fe/H] $\sim -2$ dex
which originated in smaller fragments such as Draco, Sextans and Ursa Minor. Clearly,
the measurement of even-Z abundances for stars belonging to the more massive Galactic 
satellites (such as Fornax, Sagittarius, and Leo I) are urgently needed to test 
these models.

Alternatively, it may be that the nucleosynthetic
history of the proto-Galactic fragments was affected by the building process
in such a way as to make them different than the dSph galaxies that survive today.
Still another possibility is that the old, gas-poor dSph galaxies sampled here 
comprised only a small fraction of the actual proto-Galactic fragments, and that the 
dSph galaxies with younger populations ($i.e.$, those which had more gas for 
subsequent star formation) show even-Z abundance patterns which more closely resemble
those of Galactic halo field stars. Again, element abundances
for stars belonging to an expanded sample of dwarf galaxies are required to test
these possibilities.

SBS reported that one of their giants, Draco 473, exhibits the deep mixing 
abundance pattern seen in many globular clusters, with Na enhanced and
O and Mg depleted (see Kraft 1994 and Pinsonneault 1997 for recent reviews).  
With the expanded dSph sample presented here we now reinterpret the
abundance pattern of Draco 473 as having its even-Z elements (including
O and Mg) not enhanced in comparison to Milky Way halo field stars and
probably not due to a deep mixing pattern as originally suggested.

Majewski
et al. (2000) have proposed that $\omega$ Cen may be the nucleus of
a dwarf galaxy which has been tidally stripped by the Milky Way.  If true,
a comparison of the abundances in this dwarf galaxy core to the abundances
found in the low luminosity dwarf galaxies sampled here might reveal
more clues to the formation of the halo.
The dSph galaxies studied here exhibit heavy element abundance patterns which are 
quite different from those of RGB stars in $\omega$ Cen: Smith et al. (2000) find
a large enhancement of s-process elements relative to r-process elements 
with increasing metallicity. They attribute this s-process
domination to AGB nucleosynthesis enrichment
by 1.5-3 $M_\odot$ stars. None of the dSph galaxies in our sample show 
evidence for such AGB nucleosynthesis enrichment. Thus, if
$\omega$ Cen is indeed the remains of a disrupted dwarf galaxy, then 
it must have been one different from Draco, Sextans or Ursa Minor. 
Such a situation would not be entirely surprising given the relative luminosities
of the dSph galaxies and $\omega$ Cen: $i.e.$, the latter has roughly the same
luminosity as the three dSph galaxies {\it combined}. If
$\omega$ Cen is truly the surviving nucleus a dwarf galaxy which
has been tidally stripped by the Milky Way, then more massive systems
such as Fornax or Sagittarius may be better analogs for the putative galaxy.
Once again, high-resolution abundance analysis for stars belonging to the
more massive Galactic satellites are needed to test this possibility.

As noted previously, two of the stars in our sample show large
overabundances of the heavy elements with respect to iron and an
r-process heavy element abundance pattern. Such stars are known in 
the Milky Way halo (see Westin et al. 2000; Norris et al. 1997; 
Cowan et al. 1995), but they are fairly rare.  To find two such stars in 
our small sample suggests that these stars are much more common in
dSph environments.

The underabundance of Y in our dSph sample is not easily understood, mainly 
because Y abundances in the Milky Way nucleosynthesis are themselves poorly
understood. We offer two possible solutions and suggest that higher
S/N spectra with wider wavelength coverage is required to further
understand the first s-process peak.  The [Ba/Eu] abundances in Figure 8 
suggest that the dSph sample is r-process dominated, at least among the
second s-process peak elements. If the first s-process peak were created
in a separate site from the second s-process peak, then we might expect
to find variations between the various nucleosynthetic sites.
Wasserburg et al. (1996) suggest that different r-process sites might
exist for low- and high-Z heavy elements, a hypothesis which has some observational
support ($e.g.$, Sneden et al. 2000). In an attempt to explain the deep mixing
abundance profile found in globular cluster stars, Cavallo \& Nagar (2000) suggested that
intermediate-mass AGB stars ($M > 4 M_\odot$) could produce $^{27}$Al from magnesium.
These AGB stars will produce first s-process peak elements but no second s-process
peak elements (Denissenkov et al. 1998; Boothroyd \& Sackmann 1999).   
If either of the above processes contributes to the Galactic halo [Y/Fe] abundance 
pattern, but not to that of dSph stars (or did not contribute to the gas which
made up subsequent generations of stars), then the two samples could show
different [Y/Fe] ratios. In principal, the contribution of AGB stars could be
tested carefully by searching for age spreads within these galaxies.

\subsection{DLA Comparison}

In addition to the samples of dSph and halo stars, Figures 3-8 show element abundances 
for a sample of DLA systems with low dust contents. Given the complexities 
of deriving the intrinsic ($i.e.$, undepleted) element abundances in such systems 
(see, $e.g.$, Vladilo 1998), we show only the observed abundances in these systems, 
all of which show relatively low levels of dust depletion.
For the elements Si, Cr, Mn, and Ni there is agreement among the abundances
of the dSph, Milky Way and DLA samples. As is well known, the Zn abundances of the
DLA systems are systematically larger than those of the sample of Galactic halo and
disk stars; the measured Zn abundances for the dSph stars are smaller still,
at least for the more metal-rich dSph stars.  The Zn overabundance in DLA systems 
has sometimes been attributed to depletion of Fe by dust grains ($e.g.$, Pettini et 
al. 1994; 1999; Kulkarni et al. 1997), although others contend that the usual Zn 
abundances of the DLA systems may be intrinsic to their stellar nucleosynthesis (Lu et al. 1996).

Obviously, grain depletion is not an issue for our measured dSph abundances. We find
significantly different [Zn/Fe] ratios than in the Galactic halo, suggesting that the
mechanism responsible for the production of the Zn in these galaxies was either 
missing, or was somehow kept from contributing to the enrichment of subsequent
generations of stars. The former possibility would suggest that old, low-mass 
dSph galaxies such as Draco, Ursa Minor and Sextans are not the surviving
end-products of high-redshift DLA systems. Alternatively, the latter possibility 
would indicate that the Zn-rich gas present in these galaxies would have to
be removed during a quiescent star formation period, or expelled by an energetic 
source.

Unfortunately, little is known about the nucleosynthetic site of Zn.
Classified as an iron peak element, it is largely ignored in analyses of 
neutron capture processes.  An exception is Burris et al. (2000),
who report a solar system r-process fraction of 0.66 for Zn. In light
of suggestions that different r-process sites might exist for low- and 
high-Z heavy elements, a more extensive study of Zn in relation to the 
first s-process peak elements might provide some useful insights.

\section{Summary}

High-resolution spectra for red giant stars belonging to the Draco,
Sextans and Ursa Minor dSph galaxies, and the distant Galactic globular
cluster NGC 2419, have been obtained with HIRES on the Keck I telescope.
Using these spectra, we have measured element abundances ratios for more 
than 20 elements. This is the first investigation of the abundance
patterns of stars belonging to Sextans, Ursa Minor and NGC 2419, and
constitutes a factor of four increase the number of dSph stars having
abundances measured from high-resolution spectra.

From a single red giant in NGC 2419, we find a metallicity of [Fe/H] = 
$-2.32\pm0.11$ dex, which is slightly lower than, but in acceptable agreement with, previous estimates from 
low-resolution spectroscopy (Suntzeff, Kraft \& Kinman 1988). With the possible
exception of slight enhancements in a number of heavy elements, the measured 
abundances ratios closely resemble those of red giant stars in M92: a nearby,
well-studied globular cluster of similar metallicity.

We find a remarkably uniform abundance pattern for our sample of 17 dSph stars.
This suggests that the three dSph galaxies have fairly similar 
nucleosynthetic histories and, presumably, similar initial mass
functions. All three galaxies show unmistakable evidence for a large 
range in metallicity: ${\Delta}$[Fe/H] = 1.53, 0.73 and 1.40 dex for Draco, 
Ursa Minor and Sextans, respectively. We find that the dSph stars have 
[$\alpha$/Fe] abundances which are $\simeq$ 0.2 dex {\it lower} than those of 
halo field stars in the same metallicity range: $i.e$., [$\alpha$/Fe] $\simeq$ 0.1 dex, 
compared to [$\alpha$/Fe] $\simeq$ 0.3 dex for halo stars. 
The measured [Eu/Fe] and [Ba/Fe] abundances for the dSph stars 
suggest that the chemical evolution in the dSph environments has been 
dominated by the r-process. In addition, the dSph stars show significantly 
larger [Ba/Y] abundance ratios than do halo field stars over the full range in
metallicity. These observations provide some evidence against the
notion that the Galactic halo has been assembled {\sl entirely} through
the disruption of very low-luminosity dSph galaxies like the three galaxies 
studied here. A more detailed assessment of role played by the disruption of
dwarf galaxies in the formation of the Galactic halo must
await the measurement of element abundances for stars belonging to an expanded sample
of Galactic satellites, particularly more luminous systems such as 
Sagittarius, Fornax and Leo I.

A comparison of the measured abundance ratios for the dSph stars with
those reported for DLA systems having low levels of dust depletion reveals
[Zn/Fe] abundance ratios which are nearly an order of magnitude lower
than those in the high-redshift absorbers, indicating that old, gas-poor
dSph galaxies like those studied here are probably not the low-redshift
analogs of the DLA systems.

\acknowledgments
 
This work was based on observations obtained at the W.M. Keck Observatory,
which is operated jointly by the California Institute of Technology and 
the University of California. We are grateful to the W.M. Keck Foundation 
for their vision and generosity.
We thank Tom Barlow for helpful advice on the use of MAKEE. We thank
Jon Fulbright, Andy McWilliam, John Cowan and Jim Truran for their
insightful discussions.  Thanks
also to Peter Stetson, Kyle Cudworth and Nick Suntzeff for providing
photometry for the dwarf galaxies in electronic form.
PC acknowledges support provided by the Sherman M. Fairchild Foundation.
WLWS was supported by grant AST 99-00733 from the National Science Foundation.
\clearpage

\begin{deluxetable}{crccrl}
\tablecolumns{6}
\tablenum{1}
\tablewidth{0pc}
\tablecaption{Observing Log}
\tablehead{
\colhead{Run} &
\colhead{Date} &
\colhead{Decker} &
\colhead{Binning} &
\colhead{${\lambda} / {\Delta}{\lambda}$} &
\colhead{Program Objects} 
}
\startdata
1 & 13-15/03/1999 & C5 & 1$\times$2 & 34,000 & Sextans, NGC 2419, M3, M92 \cr
2 & 15-16/07/1999 & C5 & 1$\times$2 & 34,000 & Ursa Minor, M3, M92\cr
3 & ~~~14/08/1999 & C5 & 1$\times$2 & 34,000 & Draco\cr
\enddata
\end{deluxetable}

\begin{deluxetable}{lrrcrrcr}
\tablewidth{0pt}
\tablenum{2}
\tablecaption{Properties of Program Stars}
\scriptsize
\tablehead{
\colhead{Star}&
\colhead{$V$}&
\colhead{$B-V$}&
\colhead{$E(B-V)$}&
\colhead{$T$}&
\colhead{S/N}&
\colhead{HJD}&
\colhead{$v_r$} \nl
\colhead{} &                      
\colhead{(mag)} &     
\colhead{(mag)} &     
\colhead{(mag)} &
\colhead{(sec)} &                      
\colhead{} &                      
\colhead{(+2450000.0)} &     
\colhead{(km s$^{-1}$)} 
}
\startdata
\multicolumn{8}{c}{} \nl
\multicolumn{8}{c}{{{\underbar{M92 = NGC 6341}}}} \nl
\multicolumn{8}{c}{} \nl
III-13 & 12.03 & 1.31 &0.02&  360 & 105 & 1253.10461 & -111.8$\pm$0.7 \nl
       &       &      &    &  240 &  93 & 1374.75736 & -114.1$\pm$0.3 \nl
VII-18 & 12.19 & 1.28 &0.02&  300 & 101 & 1375.75659 & -117.7$\pm$0.2 \nl
V-106  & 12.47 & 1.12 &0.02&  300 &  81 & 1375.74965 & -124.0$\pm$0.6 \nl
III-65 & 12.49 & 1.17 &0.02&  360 &  83 & 1253.11160 & -119.7$\pm$0.6 \nl
       &       &      &    &  300 &  88 & 1374.76362 & -119.0$\pm$0.3 \nl
\multicolumn{8}{c}{} \nl
\multicolumn{8}{c}{{{\underbar{M3 = NGC 5272}}}} \nl
\multicolumn{8}{c}{} \nl
III-28 & 12.81 & 1.36 &0.01&  240 &  65 & 1253.16298 & -152.8$\pm$0.2 \nl
I-21   & 13.05 & 1.36 &0.01&  300 &  65 & 1374.74650 & -145.6$\pm$0.3 \nl
IV-101 & 13.26 & 1.29 &0.01&  480 &  77 & 1375.73520 & -144.8$\pm$0.3 \nl
\multicolumn{8}{c}{} \nl
\multicolumn{8}{c}{{{\underbar{NGC 2419}}}} \nl
\multicolumn{8}{c}{} \nl
RH10   & 17.61 & 1.17 &0.06& 3600 &  20 & 1250.77596 &  -18.1$\pm$0.4 \nl
\multicolumn{8}{c}{} \nl
\multicolumn{8}{c}{{{\underbar{Draco}}}} \nl
\multicolumn{8}{c}{} \nl
11     & 17.60 & 1.12 &0.03& 3600 & 24  & 1404.82121 & -283.1$\pm$0.7 \nl
343    & 17.62 & 1.12 &0.03& 3600 & 24  & 1404.86929 & -293.1$\pm$0.8 \nl
\multicolumn{8}{c}{} \nl
\multicolumn{8}{c}{{{\underbar{Ursa Minor}}}} \nl
\multicolumn{8}{c}{} \nl
177    & 16.90 & 1.29 &0.03& 3600 &  36 & 1374.79308 & -234.5$\pm$0.3 \nl
297    & 16.91 & 1.57 &0.03& 3600 &  36 & 1374.88445 & -235.9$\pm$0.3 \nl
K      & 16.98 & 1.35 &0.03& 3600 &  34 & 1375.82789 & -246.1$\pm$0.6 \nl
O      & 17.03 & 1.31 &0.03& 3600 &  34 & 1375.87494 & -250.5$\pm$1.1 \nl
199    & 17.15 & 1.38 &0.03& 3600 &  33 & 1374.83949 & -249.2$\pm$0.3 \nl
168    & 17.88 & 0.95 &0.03& 3600 &  19 & 1375.78264 & -232.9$\pm$0.5 \nl
\multicolumn{8}{c}{} \nl
\multicolumn{8}{c}{{{\underbar{Sextans}}}} \nl
\multicolumn{8}{c}{} \nl
S35    & 17.30 & 1.41 &0.05& 3600 &  27 & 1250.92386 &  218.2$\pm$0.3 \nl
S56    & 17.37 & 1.43 &0.05& 3600 &  26 & 1250.82948 &  226.9$\pm$0.4 \nl
S49    & 17.59 & 1.15 &0.05& 3600 &  17 & 1250.87628 &  230.9$\pm$0.6 \nl
S58    & 17.69 & 1.17 &0.05& 3600 &  21 & 1251.77326 &  219.8$\pm$0.4 \nl
S36    & 17.96 & 1.10 &0.05& 3600 &  13 & 1252.77161 &  222.9$\pm$0.5 \nl
\enddata
\noindent\tt{}{STAR IDENTIFICATIONS:
Draco -- Baade \& Swope (1961);
Ursa Minor -- Van Agt (1967);
Sextans -- Suntzeff et al. (1993).}
\noindent\tt{}{REFERENCES FOR PHOTOMETRY:
M92 -- Sandage \& Walker (1966), Sandage (1970);
M3 -- Johnson \& Sandage (1956), Sandage (1953);
NGC 2419 -- Racine \& Harris (1975);
Draco -- Stetson (2000, private communication);
Ursa Minor -- Cudworth (2000, private communication);
Sextans -- Suntzeff et al. (1993).}
\end{deluxetable}

\begin{deluxetable}{crrrrrrrrrrrrrrrr}
\tablewidth{0pt}
\tablenum{3}
\tablecaption{Measured Equivalent Widths and Adopted Line Parameters}
\scriptsize
\tablehead{
\multicolumn{3}{c}{} &
\multicolumn{1}{c}{{{NGC 2419}}} &
\multicolumn{2}{c}{{{Draco}}} &
\multicolumn{6}{c}{{{Ursa Minor}}} &
\multicolumn{5}{c}{{{Sextans}}} \nl
\multicolumn{3}{c}{} &
\multicolumn{1}{c}{-----------} &
\multicolumn{2}{c}{--------------} &
\multicolumn{6}{c}{--------------------------------------------------} &
\multicolumn{5}{c}{------------------------------------------------} \nl
\colhead{Line} &
\colhead{EP} &
\colhead{${\log{gf}}$} &
\colhead{RH10} &
\colhead{11} &
\colhead{343} &
\colhead{177} &
\colhead{297} &
\colhead{K} &
\colhead{O} &
\colhead{199} &
\colhead{168} &
\colhead{S35} &
\colhead{S56} &
\colhead{S49} &
\colhead{S58} &
\colhead{S36} \nl
\colhead{(\AA)} &
\colhead{(eV)} &
\colhead{} &
\colhead{} &
\colhead{} &
\colhead{} &
\colhead{} &
\colhead{} &
\colhead{} &
\colhead{} &
\colhead{} &
\colhead{} &
\colhead{} &
\colhead{} &
\colhead{} &
\colhead{} &
\colhead{}
} 
\startdata
\multicolumn{17}{c}{}\nl
\multicolumn{17}{c}{Fe I}\nl
\multicolumn{17}{c}{}\nl
  5006.12 & 2.83  &  -0.628  &  135&  144&  135&  130&  183&  131&  136&  167&  121&  149&  155&   92&  167&  117 \nl
  5083.35 & 0.96  &  -2.862  &  169&  162&  150&  150&  223&     &  160&     &  120&  180&  179&   98&  155&  170 \nl
  5150.85 & 0.99  &  -3.030  &  143&  152&  137&  150&  215&     &  137&     &  116&  169&  154&   99&  178&  145 \nl
  5171.61 & 1.48  &  -1.751  &  163&  165&  168&  178&     &  149&  177&     &  144&     &     &  131&     &  160 \nl
  5216.28 & 1.61  &  -2.102  &  133&  144&  142&  144&     &  136&  154&     &   96&  158&  145&  106&  142&      \nl
  5217.30 & 3.21  &  -1.270  &   74&   95&   82&   82&  125&   69&   82&  112&   47&   95&   87&     &   92&   39 \nl
  5232.95 & 2.94  &  -0.067  &  145&  170&  156&  167&  228&  138&  146&     &  120&     &  170&  102&  160&  158 \nl
  5250.21 & 0.12  &  -4.700  &  119&  117&  122&  133&     &  100&  136&     &     &  159&  162&     &  141&   93 \nl
  5253.02 & 2.28  &  -3.810  &     &     &     &     &   55&     &   18&   46&     &     &   25&     &   30&      \nl
  5307.37 & 1.61  &  -2.812  &  100&  107&  100&  118&  162&   73&  106&  138&   68&  113&  141&   50&  130&   85 \nl
  5324.19 & 3.21  &  -0.100  &  118&  147&  124&  136&     &  115&  132&     &   85&  135&  151&   80&  168&  122 \nl
  5501.48 & 0.96  &  -3.050  &  160&  149&  153&  166&     &  165&  169&     &  100&     &     &  101&     &  136 \nl
  5506.79 & 0.99  &  -2.790  &  163&     &  134&  180&     &  150&  178&     &  140&     &     &  132&     &  148 \nl
  5956.70 & 0.86  &  -4.570  &   75&   87&   67&   82&  159&   46&   87&  134&   31&  100&  114&   23&  129&   67 \nl
  5976.79 & 3.94  &  -1.290  &     &   41&   31&   28&   57&     &     &   46&     &     &     &     &     &      \nl
  6027.06 & 4.07  &  -1.180  &     &   32&     &   30&   53&     &     &   68&     &   18&     &     &     &      \nl
  6056.01 & 4.73  &  -0.450  &     &   29&     &     &   50&     &   20&   40&   16&     &     &     &     &      \nl
  6078.50 & 4.79  &  -0.370  &     &     &     &   21&   46&     &   19&   40&     &     &     &     &     &      \nl
  6079.01 & 4.65  &  -0.950  &     &     &     &     &     &     &     &   25&     &     &     &     &     &      \nl
  6082.72 & 2.22  &  -3.590  &     &     &     &   25&   86&     &   33&   73&     &     &     &     &     &      \nl
  6120.26 & 0.91  &  -5.940  &     &     &     &     &   40&     &     &   33&     &     &     &     &     &      \nl
  6136.62 & 2.45  &  -1.500  &  133&  140&  133&  141&  220&  118&  141&     &   87&  149&  156&   64&  158&  148 \nl
  6137.70 & 2.59  &  -1.366  &  132&  125&  106&  123&  211&  113&  134&     &   81&  147&  157&   54&  148&  142 \nl
  6151.62 & 2.18  &  -3.370  &     &   46&   34&   39&  106&   19&   50&   83&   17&   60&   67&     &   83&      \nl
  6157.73 & 4.07  &  -1.260  &     &   38&     &   24&   73&     &   27&     &     &   34&     &     &   37&      \nl
  6165.36 & 4.14  &  -1.470  &     &     &     &     &   24&     &     &   34&     &     &     &     &     &      \nl
  6173.34 & 2.22  &  -2.850  &   50&   82&   55&   72&  130&   46&   73&  115&   30&   83&   80&     &   97&   47 \nl
  6187.99 & 3.94  &  -1.580  &     &     &     &     &   38&     &     &   48&     &     &     &     &     &      \nl
  6213.43 & 2.22  &  -2.660  &   72&     &     &  100&     &     &   98&  123&     &  119&  104&     &  130&   89 \nl
  6229.23 & 2.84  &  -2.900  &     &   20&   21&   17&   46&     &     &   51&     &   29&   34&     &   35&      \nl
  6230.74 & 2.56  &  -1.276  &  138&  163&  144&  148&  227&  138&  155&     &  105&  167&  165&   87&  148&  124 \nl
  6240.66 & 2.22  &  -3.230  &     &   50&   24&   47&  107&     &   40&   77&     &   69&   65&     &   59&      \nl
  6252.57 & 2.40  &  -1.757  &  137&  121&  122&  135&  172&  108&  139&  167&   99&  144&  155&   62&  138&  125 \nl
  6290.97 & 4.73  &  -0.760  &     &     &     &     &   32&     &     &     &     &     &     &     &     &      \nl
  6297.80 & 2.22  &  -2.740  &     &   66&   87&  105&  143&     &   90&  120&   57&     &     &     &     &      \nl
  6353.84 & 0.91  &  -6.430  &     &     &     &     &   11&     &     &   11&     &     &     &     &     &      \nl
  6355.04 & 2.84  &  -2.290  &     &   62&   48&   51&  123&     &   58&   94&     &     &     &     &     &      \nl
  6380.75 & 4.19  &  -1.500  &     &     &     &     &   54&     &     &   42&     &     &     &     &     &      \nl
  6392.54 & 2.28  &  -3.950  &     &     &   16&     &   50&     &     &   38&     &     &     &     &     &      \nl
  6393.61 & 2.43  &  -1.630  &     &  147&  131&  129&     &  155&  138&  179&   82&     &     &     &     &      \nl
  6498.94 & 0.96  &  -4.690  &     &   86&   62&   64&  158&   48&   80&  113&     &     &     &     &     &      \nl
  6518.37 & 2.83  &  -2.460  &     &   56&   47&   44&   89&   25&   38&   81&     &     &     &     &     &      \nl
  6574.23 & 0.99  &  -5.020  &     &   50&   49&   53&  114&   39&   48&   94&     &     &     &     &     &      \nl
  6581.22 & 1.48  &  -4.680  &     &     &   33&   21&   77&     &   21&   59&     &     &     &     &     &      \nl
  6593.88 & 2.43  &  -2.390  &     &   82&   80&   87&  143&   80&   92&  138&     &     &     &     &     &      \nl
  6608.03 & 2.28  &  -3.940  &     &     &     &     &   40&     &     &   26&     &     &     &     &     &      \nl
  6609.12 & 2.56  &  -2.660  &     &   59&   44&   43&  119&   44&   69&  106&     &     &     &     &     &      \nl
\multicolumn{17}{c}{}\nl
\multicolumn{17}{c}{Fe II} \nl                                                                                   
\multicolumn{17}{c}{}\nl
  5234.63 & 3.22  &  -2.118  &   66&   87&   84&   87&   75&   88&   85&   95&   65&   89&   88&   75&  121&   70 \nl
  5264.81 & 3.23  &  -3.210  &     &   37&   31&   23&   40&     &   33&   48&     &   30&   33&     &   45&   27 \nl
  6149.25 & 3.89  &  -2.720  &     &   30&   25&   22&   17&   23&   31&   39&     &   29&   20&     &   40&      \nl
  6369.46 & 2.89  &  -4.250  &     &     &   17&     &     &     &     &   29&     &     &     &     &     &      \nl
  6456.39 & 3.90  &  -2.080  &     &   52&   49&   46&   58&   53&   50&   71&   28&     &     &     &     &      \nl
  6516.08 & 2.89  &  -3.450  &     &   58&   44&   43&   44&   44&   55&   59&     &     &     &     &     &      \nl
\multicolumn{17}{c}{}\nl
\multicolumn{17}{c}{O I} \nl                                                                                     
\multicolumn{17}{c}{}\nl
  6363.79 & 0.02  & -10.250  &     &$<$10&   14&$<$10&   17&$<$10&$<$10&$<$10&$<$10&     &     &     &     &      \nl
\multicolumn{17}{c}{}\nl
\multicolumn{17}{c}{Na I} \nl                                                                                    
\multicolumn{17}{c}{}\nl
  5682.65 & 2.10  &  -0.700  &$<$19&$<$15&$<$18&$<$15&   73&$<$18&$<$15&$<$12&$<$20&$<$15& $<$8&$<$22&$<$20&$<$19 \nl
  5688.21 & 2.10  &  -0.370  &$<$33&$<$20&$<$20&   15&   88&$<$20&$<$15&   27&$<$30&   25&   25&$<$20&$<$30&$<$20 \nl
  6154.23 & 2.10  &  -1.560  &     &     &     &     &$<$13&     &     &     &     &     &     &     &     &      \nl
  6160.75 & 2.10  &  -1.260  &     &     &     &     &   29&     &     &     &     &     &     &     &     &      \nl
\multicolumn{17}{c}{}\nl
\multicolumn{17}{c}{Mg I} \nl                                                                                    
\multicolumn{17}{c}{}\nl
  4703.00 & 4.33  &  -0.520  &   99&  123&  120&  119&  177&  124&  129&  128&  102&  118&  108&   88&   94&      \nl
  5172.70 & 2.71  &  -0.390  &  290&  298&  293&  298&  457&  280&  330&  368&  259&  310&  349&  235&  315&  264 \nl
  5183.27 & 2.70  &  -0.170  &  386&  372&  327&  376&  660&  400&  400&  505&  305&  453&  438&  228&  362&  280 \nl
\multicolumn{17}{c}{}\nl
\multicolumn{17}{c}{Al I} \nl                                                                                    
\multicolumn{17}{c}{}\nl
  6696.03 & 3.14  &  -1.570  &     &$<$20&$<$10&$<$10&$<$10&$<$10&$<$14&$<$10&$<$10&     &     &     &     &      \nl
\multicolumn{17}{c}{}\nl
\multicolumn{17}{c}{Si I} \nl                                                                                    
\multicolumn{17}{c}{}\nl
  5645.66 & 4.91  &  -2.140  &     &     &     &     &$<$15&     &     &     &     &     &     &     &     &      \nl
  5665.60 & 4.90  &  -2.040  &$<$10&     &$<$20&     &$<$13&$<$25&$<$15&$<$15&$<$18&$<$15&     &$<$20&$<$20&$<$22 \nl
  5684.52 & 4.93  &  -1.650  &$<$19&$<$15&$<$16&   16&   34&   23&   20&$<$12&$<$15&   17&$<$16&$<$20&$<$30&$<$27 \nl
  5772.26 & 5.06  &  -1.750  &$<$15&$<$15&$<$20&$<$12&$<$15&$<$18&$<$20&$<$12&$<$20&     &$<$20&$<$20&$<$20&$<$30 \nl
  6145.02 & 5.61  &  -1.370  &     &     &$<$15&   16&$<$12&$<$12&     &   12&     &     &     &     &$<$15&      \nl
  6243.82 & 5.61  &  -1.270  &     &$<$12&$<$12&     &     &     &     &$<$12&     &$<$15&$<$12&     &$<$18&      \nl
  6244.48 & 5.61  &  -1.270  &     &$<$15&     &$<$12&   18&     &$<$18&   15&$<$20&   12&$<$15&$<$20&$<$20&$<$22 \nl
\multicolumn{17}{c}{}\nl
\multicolumn{17}{c}{Ca I} \nl                                                                                    
\multicolumn{17}{c}{}\nl
  6122.23 & 1.89  &  -0.320  &  124&  138&  120&  138&  188&  121&  139&  165&  107&  149&  142&   58&  143&  130 \nl
  6161.30 & 2.52  &  -1.270  &     &   29&     &   22&   53&     &   22&   64&     &   26&     &     &   35&   30 \nl
  6166.44 & 2.52  &  -1.140  &     &   42&     &   22&   63&     &     &   45&     &   28&   40&     &     &   30 \nl
  6169.04 & 2.52  &  -0.800  &   27&   63&   39&   42&   85&   45&   40&   64&     &   45&   45&     &   43&   28 \nl
  6169.56 & 2.52  &  -0.480  &   36&   75&   47&   65&   95&   52&   55&   81&   21&   63&   72&     &   72&   50 \nl
  6455.60 & 2.52  &  -1.290  &     &   22&     &   17&   58&     &   15&   45&     &     &     &     &     &      \nl
  6499.65 & 2.52  &  -0.820  &     &   44&   42&   48&   88&   58&   47&   60&     &     &     &     &     &      \nl
\multicolumn{17}{c}{}\nl
\multicolumn{17}{c}{Ti I} \nl                                                                                    
\multicolumn{17}{c}{}\nl
  4548.76 & 0.83  &  -0.290  &   50&     &     &   69&     &     &   94&  106&     &   98&  104&     &     &   44 \nl
  4555.49 & 0.85  &  -0.430  &   60&     &     &   63&  113&   62&   64&   85&   44&   67&   98&     &   61&      \nl
  4623.10 & 1.74  &   0.166  &   31&   48&     &   25&   75&     &   39&   50&     &     &     &     &     &      \nl
  4645.19 & 1.74  &  -0.501  &    5&     &     &     &   58&     &     &     &     &     &   35&     &     &      \nl
  4656.47 & 0.00  &  -1.280  &  103&   83&   85&   87&     &   51&   76&  128&   45&  120&  107&     &   89&   40 \nl
  4840.87 & 0.90  &  -0.450  &   76&   54&   39&   73&  140&   34&   61&   87&   33&   82&   93&     &   74&      \nl
  4913.62 & 1.87  &   0.216  &     &   35&     &   19&   71&     &     &   31&     &     &   42&     &   29&      \nl
  4997.10 & 0.00  &  -2.060  &   46&     &     &   64&  151&     &   69&   86&     &   52&   58&     &   47&   60 \nl
  5014.24 & 0.81  &   0.910  &  196&  167&  139&  162&  273&  135&  178&  220&  100&  200&  188&   60&  138&  113 \nl
  5016.16 & 0.85  &  -0.510  &   54&   74&   53&   80&  130&   44&   77&  109&   36&   76&   89&     &   62&   47 \nl
  5064.65 & 0.05  &  -0.930  &  114&  133&   96&  113&  210&   94&  124&     &   67&  131&  128&   31&  111&  110 \nl
  5113.44 & 1.44  &  -0.727  &     &     &     &   18&   62&     &     &   25&     &     &     &     &     &      \nl
  5145.47 & 1.46  &  -0.518  &     &     &     &   29&   64&     &     &   38&     &     &     &     &     &      \nl
  5210.39 & 0.05  &  -0.580  &  112&  118&  118&  136&  221&  115&  148&  164&   75&  136&  176&   30&  133&   84 \nl
  5978.54 & 1.87  &  -0.440  &     &     &     &   14&   34&     &   26&   30&     &     &     &     &     &      \nl
\multicolumn{17}{c}{}\nl
\multicolumn{17}{c}{Ti II} \nl                                                                                   
\multicolumn{17}{c}{}\nl
  4583.41 & 1.16  &  -2.870  &     &   55&     &   43&   73&   35&   70&   78&     &   48&   42&     &   40&      \nl
  4636.32 & 1.16  &  -3.230  &     &   48&   52&   51&   58&     &   35&     &     &   25&     &     &   45&      \nl
  4657.20 & 1.24  &  -2.320  &   53&  108&   75&     &   97&   67&   83&   84&   50&   75&   79&     &   39&      \nl
  4708.66 & 1.24  &  -2.370  &   78&   77&   65&   73&   85&     &   80&   87&     &   76&   88&     &   62&      \nl
  4719.52 & 1.24  &  -3.280  &     &   20&   37&   30&   40&     &   18&   50&     &     &     &     &     &      \nl
  4798.53 & 1.08  &  -2.670  &     &   77&   73&   85&  103&     &   87&  101&     &   82&     &     &     &   38 \nl
  5154.07 & 1.57  &  -1.520  &   90&   96&   82&  104&   95&  100&  100&  132&   60&  105&  102&   40&   95&   75 \nl
  5226.55 & 1.57  &  -1.000  &  105&  115&   91&  119&  152&  117&  125&  135&  105&  120&  125&   74&  117&  100 \nl
  5381.01 & 1.57  &  -1.780  &   79&   80&   72&   84&  135&   76&   84&  120&   45&  123&   90&   45&   81&   65 \nl
  5418.77 & 1.58  &  -2.110  &   74&   87&   57&   70&   88&   65&   86&   69&   47&   69&   52&   19&   63&   46 \nl
\multicolumn{17}{c}{}\nl
\multicolumn{17}{c}{V I} \nl                                                                                     
\multicolumn{17}{c}{}\nl
  6216.37 & 0.28  &  -1.270  &     &     &   18&     &   86&     &   18&   35&     &   24&   26&     &     &      \nl
  6224.51 & 0.29  &  -1.820  &     &     &     &     &   29&     &     &   23&     &     &     &     &     &      \nl
  6233.20 & 0.28  &  -2.000  &     &     &     &     &     &     &     &     &     &     &   11&     &     &      \nl
  6274.66 & 0.27  &  -1.690  &     &     &     &     &   57&     &     &     &     &     &     &     &     &      \nl
  6285.16 & 0.28  &  -1.560  &     &     &     &     &   75&     &     &     &     &     &     &     &     &      \nl
  6292.82 & 0.29  &  -1.520  &     &     &     &     &     &     &   25&   39&     &     &     &     &     &      \nl
\multicolumn{17}{c}{}\nl
\multicolumn{17}{c}{Cr I} \nl                                                                                    
\multicolumn{17}{c}{}\nl
  5409.80 & 1.03  &  -0.720  &  111&  136&  100&  125&  221&  119&  124&  177&   59&  139&  148&   40&  138&  100 \nl
\multicolumn{17}{c}{}\nl
\multicolumn{17}{c}{Mn I} \nl                                                                                    
\multicolumn{17}{c}{}\nl
  5407.42 & 2.14  &  -1.743  &     &     &     &   16&   52&     &     &   48&     &     &     &     &   25&$<$20 \nl
  5420.36 & 2.14  &  -1.460  &   13&$<$18&     &   23&   89&$<$20&   30&   66&$<$20&   35&   31&     &     &      \nl
  5432.55 & 0.00  &  -3.795  &   30&   45&   38&   34&  165&   26&   51&     &   21&   75&   50&$<$20&   55&$<$20 \nl
  5516.77 & 2.18  &  -1.847  &     &     &     &     &   50&     &     &   23&     &     &     &     &     &      \nl
  6013.51 & 3.07  &  -0.252  &$<$12&$<$15&   16&   14&   54&$<$25&$<$25&   51&$<$23&   20&$<$18&$<$20&   37&$<$20 \nl
  6021.82 & 3.08  &   0.035  &   42&   46&$<$18&   31&   93&   30&   49&   64&$<$24&   35&   40&$<$27&   53&$<$25 \nl
\multicolumn{17}{c}{}\nl
\multicolumn{17}{c}{Co I} \nl                                                                                    
\multicolumn{17}{c}{}\nl
  5483.34 & 1.71  &  -1.488  &$<$23&   33&   25&   54&  130&$<$23&   34&   61&$<$18&   39&   49&$<$30&   22&$<$23 \nl
  5647.23 & 2.28  &  -1.560  &     &     &     &$<$15&   18&     &     &   15&     &     &     &     &     &      \nl
  6454.99 & 3.63  &  -0.250  &     &     &     &     &   19&     &     &     &     &     &     &     &     &      \nl
\multicolumn{17}{c}{}\nl
\multicolumn{17}{c}{Ni I} \nl                                                                                    
\multicolumn{17}{c}{}\nl
  5476.92 & 1.83  &  -0.890  &  142&  138&  137&  148&  205&  134&  148&  200&  110&  164&  158&  116&  147&  123 \nl
  6176.82 & 4.09  &  -0.430  &     &     &     &   20&   25&     &     &   22&     &     &     &     &     &      \nl
  6177.25 & 1.83  &  -3.500  &     &   20&     &     &     &     &     &   15&     &     &     &     &     &      \nl
\multicolumn{17}{c}{}\nl
\multicolumn{17}{c}{Cu I} \nl                                                                                    
\multicolumn{17}{c}{}\nl
  5782.13 & 1.64  &  -1.781  &$<$14&$<$30&$<$20&$<$20&   56&$<$18&$<$18&$<$30&$<$20&$<$18&$<$20&$<$40&$<$30&$<$40 \nl
\multicolumn{17}{c}{}\nl
\multicolumn{17}{c}{Zn I} \nl                                                                                    
\multicolumn{17}{c}{}\nl
  4722.16 & 4.03  &  -0.390  &   48&   45&   45&   41&   60&     &   67&   63&   25&   36&   26&     &     &      \nl
  4810.54 & 4.08  &  -0.170  &   12&   41&   23&   36&   38&   38&   59&   44&   37&   32&   44&$<$29&   42&   35 \nl
\multicolumn{17}{c}{}\nl
\multicolumn{17}{c}{Y II} \nl                                                                                    
\multicolumn{17}{c}{}\nl
  4883.69 & 1.08  &   0.070  &   77&   62&   47&   61&  108&   90&   52&  130&   34&   76&   65&$<$20&   61&   65 \nl
  4900.11 & 1.03  &  -0.090  &   82&   50&   70&   54&     &   81&   60&  128&   50&   90&   74&     &   70&  100 \nl
  5087.43 & 1.08  &  -0.170  &   55&   37&     &   43&   98&   76&   38&  114&   31&     &     &     &     &      \nl
  5200.42 & 0.99  &  -0.570  &   40&   39&   40&   43&   90&   52&   38&  110&   29&   74&   46&$<$23&   27&   69 \nl
\multicolumn{17}{c}{}\nl
\multicolumn{17}{c}{Ba II} \nl                                                                                   
\multicolumn{17}{c}{}\nl
  5853.69 & 0.60  &  -1.010  &   82&   99&   79&   82&  142&  180&   77&  177&   62&  165&  127&   20&  120&  100 \nl
  6141.73 & 0.70  &  -0.077  &  128&  150&  150&  144&  207&  269&  135&  249&  142&  204&  172&   52&  165&  175 \nl
  6496.91 & 0.60  &  -0.380  &     &  163&  163&  143&  212&  265&  133&  251&  115&     &     &     &     &      \nl
\multicolumn{17}{c}{}\nl
\multicolumn{17}{c}{Ce II} \nl                                                                                   
\multicolumn{17}{c}{}\nl
  4562.37 & 0.48  &   0.500  &   35&   50&   29&   30&   73&  104&   15&  134&   23&  102&   44&$<$40&   40&$<$65 \nl
  4628.16 & 0.52  &   0.390  &   38&   55&   25&   30&   60&  122&   15&  121&   22&  105&   50&$<$50&   40&$<$65 \nl
\multicolumn{17}{c}{}\nl
\multicolumn{17}{c}{Nd II} \nl                                                                                   
\multicolumn{17}{c}{}\nl
  5249.59 & 0.98  &   0.217  &   19&   43&   16&   27&   78&   87&   20&  110&   23&   71&   56&$<$30&   45&   44 \nl
  5319.82 & 0.55  &  -0.194  &   37&   55&   30&   34&   90&  111&   34&  133&   19&  101&   48&$<$25&   45&   42 \nl
\multicolumn{17}{c}{}\nl
\multicolumn{17}{c}{Sm II} \nl                                                                                   
\multicolumn{17}{c}{}\nl
  4642.23 & 0.38  &  -0.520  &$<$35&   33&   20&   35&   91&   90&   30&  121&$<$20&   95&   59&$<$25&   50&$<$65 \nl
\multicolumn{17}{c}{}\nl
\multicolumn{17}{c}{Eu II} \nl                                                                                   
\multicolumn{17}{c}{}\nl
  6645.13 & 1.37  &   0.200  &     &   27&$<$20&   15&   49&   50&   15&  107&   36&     &     &     &     &      \nl
\enddata
\end{deluxetable}

\begin{deluxetable}{lcccrrrrrrrr}
\tablewidth{0pt}
\scriptsize
\tablenum{4}
\tablecaption{Element Abundances for Program Stars}
\tablehead{
\colhead{Star}&
\colhead{T$_{\rm eff}$}&
\colhead{${\log{g}}$}&
\colhead{$v_t$}&
\colhead{[Fe/H]}&
\colhead{[O/Fe]}&
\colhead{[Na/Fe]}&
\colhead{[Mg/Fe]}&
\colhead{[Al/Fe]}&
\colhead{[Si/Fe]}&
\colhead{[Ca/Fe]}&
\colhead{[Ti/Fe]} \nl
\colhead{}&
\colhead{(K)}&
\colhead{(dex)}&
\colhead{(km s$^{-1}$)}&
\colhead{(dex)}&
\colhead{(dex)}&
\colhead{(dex)}&
\colhead{(dex)}&
\colhead{(dex)}&
\colhead{(dex)}&
\colhead{(dex)}&
\colhead{(dex)}
}
\startdata
\multicolumn{12}{c}{} \nl
\multicolumn{12}{c}{{{\underbar{M92 = NGC 6341}}}} \nl
\multicolumn{12}{c}{} \nl
III-13 & 4175 & 0.20 & 2.10 & -2.28$\pm$0.11 & 0.13$\pm$0.15 & 0.48$\pm$0.10 & 0.64$\pm$0.16 & 0.90$\pm$0.15 & 0.47$\pm$0.15 & 0.30$\pm$0.08 & 0.39$\pm$0.07 \nl
VII-18 & 4225 & 0.00 & 2.15 & -2.27$\pm$0.11 & $<$0.15       & 0.38$\pm$0.15 & 0.37$\pm$0.15 & 1.25$\pm$0.15 & 0.75$\pm$0.18 & 0.29$\pm$0.08 & 0.32$\pm$0.07 \nl
V-106  & 4375 & 0.70 & 2.00 & -2.33$\pm$0.11 & $<$0.36       &-0.21$\pm$0.15 & 0.48$\pm$0.15 & $<$0.90       & 0.47$\pm$0.15 & 0.26$\pm$0.08 & 0.24$\pm$0.07 \nl
III-65 & 4325 & 0.50 & 2.10 & -2.30$\pm$0.11 & 0.23$\pm$0.15 & 0.31$\pm$0.14 & 0.38$\pm$0.18 & 1.01$\pm$0.18 & 0.62$\pm$0.20 & 0.27$\pm$0.07 & 0.15$\pm$0.06 \nl
\multicolumn{12}{c}{} \nl
\multicolumn{12}{c}{{{\underbar{M3 = NGC 5272}}}} \nl
\multicolumn{12}{c}{} \nl
III-28 & 4175 & 0.70 & 1.90 & -1.70$\pm$0.11 & 0.37$\pm$0.14 &-0.23$\pm$0.09 & 0.41$\pm$0.15 &-0.06$\pm$0.20 & 0.31$\pm$0.12 & 0.28$\pm$0.08 & 0.37$\pm$0.07 \nl
I-21   & 4200 & 0.70 & 1.70 & -1.44$\pm$0.11 & 0.21$\pm$0.14 &-0.26$\pm$0.09 & 0.30$\pm$0.15 & 0.24$\pm$0.14 & 0.15$\pm$0.14 & 0.27$\pm$0.07 & 0.32$\pm$0.06 \nl
IV-101 & 4225 & 0.80 & 1.70 & -1.44$\pm$0.11 & 0.02$\pm$0.14 & 0.20$\pm$0.09 & 0.24$\pm$0.13 & 0.85$\pm$0.12 & 0.09$\pm$0.15 & 0.28$\pm$0.08 & 0.32$\pm$0.07 \nl
\multicolumn{12}{c}{} \nl
\multicolumn{12}{c}{{{\underbar{NGC 2419}}}} \nl
\multicolumn{12}{c}{} \nl
RH10   & 4275 & 0.70 & 2.10 & -2.32$\pm$0.11 &               & $<$0.26       & 0.30$\pm$0.18 &               & $<$0.72       & 0.11$\pm$0.12 & 0.22$\pm$0.11 \nl
\multicolumn{12}{c}{} \nl
\multicolumn{12}{c}{{{\underbar{Draco}}}} \nl
\multicolumn{12}{c}{} \nl
11     & 4475 & 0.80 & 1.80 & -1.72$\pm$0.11 & $<$0.38       &$<$-0.34       & 0.07$\pm$0.15 & $<$0.71       & $<$0.30       & 0.16$\pm$0.08 & 0.09$\pm$0.07 \nl
343    & 4475 & 0.90 & 1.80 & -1.86$\pm$0.11 & 0.42$\pm$0.22 &-0.15$\pm$0.00 & 0.06$\pm$0.20 & $<$0.34       & $<$0.50       & 0.03$\pm$0.11 &-0.03$\pm$0.11 \nl
473    & 4400 & 0.90 & 1.75 & -1.44$\pm$0.07 &-0.32$\pm$0.18 &-0.04$\pm$0.09 &-0.19$\pm$0.17 &               & $<$0.14       & 0.18$\pm$0.08 &-0.18$\pm$0.25 \nl
267    & 4180 & 0.60 & 1.95 & -1.67$\pm$0.13 & 0.21$\pm$0.11 &-0.59$\pm$0.17 &-0.04$\pm$0.21 &               & $<$0.29       & 0.19$\pm$0.08 & 0.02$\pm$0.31 \nl
24     & 4290 & 0.80 & 2.00 & -2.36$\pm$0.09 & 0.38$\pm$0.18 &-0.33$\pm$0.18 & 0.26$\pm$0.16 &               &               & 0.07$\pm$0.06 &-0.04$\pm$0.18 \nl
119    & 4370 & 0.15 & 2.80 & -2.97$\pm$0.15 &               &-0.09$\pm$0.17 & 0.20$\pm$0.17 &               &               & 0.11$\pm$0.07 &-0.17$\pm$0.31 \nl
\multicolumn{12}{c}{} \nl
\multicolumn{12}{c}{{{\underbar{Ursa Minor}}}} \nl
\multicolumn{12}{c}{} \nl
177    & 4300 & 0.40 & 1.90 & -2.01$\pm$0.11 & $<$0.33       &-0.37$\pm$0.18 & 0.32$\pm$0.17 & $<$0.68       & 0.51$\pm$0.30 & 0.18$\pm$0.08 & 0.15$\pm$0.08 \nl
297    & 4075 & 0.40 & 2.30 & -1.68$\pm$0.11 & 0.18$\pm$0.22 & 0.15$\pm$0.12 & 0.24$\pm$0.22 & $<$0.21       & 0.30$\pm$0.20 & 0.02$\pm$0.08 & 0.12$\pm$0.07 \nl
K      & 4325 & 0.10 & 2.00 & -2.17$\pm$0.12 & $<$0.38       & $<$0.20       & 0.69$\pm$0.25 & $<$0.90       & $<$0.88       & 0.43$\pm$0.11 & 0.04$\pm$0.10 \nl
O      & 4325 & 0.30 & 1.80 & -1.91$\pm$0.11 & $<$0.22       &$<$-0.24       & 0.43$\pm$0.20 & $<$0.51       & 0.32$\pm$0.30 & 0.09$\pm$0.08 & 0.20$\pm$0.07 \nl
199    & 4325 & 0.30 & 1.95 & -1.45$\pm$0.11 &$<$-0.05       &$<$-0.66       & 0.02$\pm$0.17 & $<$0.01       & 0.00$\pm$0.20 &-0.01$\pm$0.09 &-0.08$\pm$0.09 \nl
168    & 4625 & 1.30 & 1.70 & -2.18$\pm$0.12 &               & $<$0.30       & 0.23$\pm$0.17 &               & $<$0.89       & 0.11$\pm$0.30 & 0.01$\pm$0.14 \nl
\multicolumn{12}{c}{} \nl
\multicolumn{12}{c}{{{\underbar{Sextans}}}} \nl
\multicolumn{12}{c}{} \nl
S35    & 4225 & 0.10 & 2.00 & -1.93$\pm$0.11 &               &-0.24$\pm$0.18 & 0.27$\pm$0.19 &               & 0.45$\pm$0.30 & 0.11$\pm$0.10 & 0.07$\pm$0.10 \nl
S56    & 4175 & 0.30 & 1.90 & -1.93$\pm$0.11 &               &-0.27$\pm$0.18 & 0.23$\pm$0.19 &               & $<$0.45       & 0.08$\pm$0.10 & 0.13$\pm$0.10 \nl
S49    & 4325 & 0.10 & 2.50 & -2.85$\pm$0.13 &               &               & 0.41$\pm$0.20 &               &               & 0.08$\pm$0.23 &-0.29$\pm$0.15 \nl
S58    & 4525 & 1.00 & 1.90 & -1.45$\pm$0.12 &               &$<$-0.43       &-0.46$\pm$0.19 &               & $<$0.30       &-0.12$\pm$0.10 &-0.35$\pm$0.10 \nl
S36    & 4425 & 1.10 & 2.10 & -2.19$\pm$0.12 &               & $<$0.15       &-0.07$\pm$0.20 &               &               & 0.34$\pm$0.11 &-0.10$\pm$0.11 \nl
\enddata
\end{deluxetable}

\clearpage

\begin{deluxetable}{lrrrrrrrrr}
\tablewidth{0pt}
\scriptsize
\tablenum{4 Cont'd}
\tablecaption{Element Abundances for Program Stars}
\tablehead{
\colhead{Star}&
\colhead{[Fe/H]}&
\colhead{[V/Fe]}&
\colhead{[Cr/Fe]}&
\colhead{[Mn/Fe]}&
\colhead{[Co/Fe]}&
\colhead{[Ni/Fe]}&
\colhead{[Cu/Fe]}&
\colhead{[Zn/Fe]}&
\colhead{[Cr/Co]} \nl
\colhead{}&
\colhead{(dex)}&
\colhead{(dex)}&
\colhead{(dex)}&
\colhead{(dex)}&
\colhead{(dex)}&
\colhead{(dex)}&
\colhead{(dex)}&
\colhead{(dex)}&
\colhead{(dex)}
}
\startdata
\multicolumn{10}{c}{} \nl
\multicolumn{10}{c}{{{\underbar{M92 = NGC 6341}}}} \nl
\multicolumn{10}{c}{} \nl
III-13 & -2.28$\pm$0.11 & 0.04$\pm$0.10 &-0.15$\pm$0.12 &-0.27$\pm$0.08 & 0.18$\pm$0.10 & 0.04$\pm$0.12 &-0.64$\pm$0.15 & 0.17$\pm$0.14 &  0.33$\pm$0.16 \nl
VII-18 & -2.27$\pm$0.11 &-0.02$\pm$0.12 &-0.10$\pm$0.12 &-0.37$\pm$0.09 & 0.08$\pm$0.10 & 0.02$\pm$0.12 &-0.65$\pm$0.15 & 0.30$\pm$0.16 &  0.18$\pm$0.16 \nl
V-106  & -2.33$\pm$0.11 & 0.43$\pm$0.20 &-0.23$\pm$0.14 &-0.28$\pm$0.14 & 0.05$\pm$0.11 & 0.13$\pm$0.13 &-0.76$\pm$0.15 & 0.14$\pm$0.12 &  0.28$\pm$0.18 \nl
III-65 & -2.30$\pm$0.11 & 0.17$\pm$0.10 &-0.27$\pm$0.11 &-0.42$\pm$0.14 & 0.15$\pm$0.15 & 0.10$\pm$0.12 &-0.78$\pm$0.15 & 0.07$\pm$0.12 &  0.42$\pm$0.19 \nl
\multicolumn{10}{c}{} \nl
\multicolumn{10}{c}{{{\underbar{M3 = NGC 5272}}}} \nl
\multicolumn{10}{c}{} \nl
III-28 & -1.70$\pm$0.11 & 0.12$\pm$0.09 &-0.01$\pm$0.13 &-0.26$\pm$0.08 &-0.04$\pm$0.20 & 0.11$\pm$0.13 &-0.37$\pm$0.13 & 0.09$\pm$0.14 & -0.03$\pm$0.24 \nl
I-21   & -1.44$\pm$0.11 & 0.08$\pm$0.09 & 0.08$\pm$0.12 &-0.14$\pm$0.10 &-0.08$\pm$0.16 &-0.04$\pm$0.12 &-0.17$\pm$0.13 &-0.02$\pm$0.15 & -0.16$\pm$0.20 \nl
IV-101 & -1.44$\pm$0.11 & 0.08$\pm$0.09 & 0.12$\pm$0.13 &-0.19$\pm$0.10 & 0.07$\pm$0.14 &-0.07$\pm$0.16 &-0.28$\pm$0.13 & 0.05$\pm$0.10 & -0.05$\pm$0.19 \nl
\multicolumn{10}{c}{} \nl
\multicolumn{10}{c}{{{\underbar{NGC 2419}}}} \nl
\multicolumn{10}{c}{} \nl
RH10   & -2.32$\pm$0.11 &               &-0.32$\pm$0.17 &-0.31$\pm$0.12 & $<$0.00       & 0.00$\pm$0.19 &$<$-0.43       & $<$0.15       &                \nl
\multicolumn{10}{c}{} \nl
\multicolumn{10}{c}{{{\underbar{Draco}}}} \nl
\multicolumn{10}{c}{} \nl
11     & -1.72$\pm$0.11 &               &-0.01$\pm$0.17 &-0.50$\pm$0.10 &-0.19$\pm$0.18 & 0.00$\pm$0.15 &$<$-0.35       &-0.21$\pm$0.15 & -0.18$\pm$0.25 \nl
343    & -1.86$\pm$0.11 &-0.02$\pm$0.18 &-0.49$\pm$0.18 &-0.56$\pm$0.10 &-0.20$\pm$0.18 &-0.06$\pm$0.21 &$<$-0.43       &-0.28$\pm$0.30 &  0.29$\pm$0.25 \nl
473    & -1.44$\pm$0.07 &-0.68$\pm$0.18 &-0.05$\pm$0.12 &-0.32$\pm$0.15 & 0.12$\pm$0.14 & 0.19$\pm$0.13 &$<$-0.10       & 0.12$\pm$0.20 &  0.17$\pm$0.18 \nl
267    & -1.67$\pm$0.13 &-0.20$\pm$0.18 &-0.05$\pm$0.13 &-0.28$\pm$0.16 & $<$0.05       & 0.36$\pm$0.14 &$<$-0.39       &$<$-0.46       &                \nl
24     & -2.36$\pm$0.09 &               & 0.04$\pm$0.11 & $<$0.05       &-0.07$\pm$0.20 & 0.31$\pm$0.16 &$<$-0.15       & 0.10$\pm$0.20 & -0.11$\pm$0.23 \nl
119    & -2.97$\pm$0.15 &               &-0.54$\pm$0.08 & $<$0.80       & $<$1.27       &-0.42$\pm$0.11 & $<$0.92       & $<$1.06       &                \nl
\multicolumn{10}{c}{} \nl
\multicolumn{10}{c}{{{\underbar{Ursa Minor}}}} \nl
\multicolumn{10}{c}{} \nl
177    & -2.01$\pm$0.11 &               &-0.23$\pm$0.17 &-0.37$\pm$0.10 & 0.26$\pm$0.13 & 0.08$\pm$0.17 &$<$-0.50       &-0.01$\pm$0.15 &  0.49$\pm$0.21 \nl
297    & -1.68$\pm$0.11 &-0.13$\pm$0.10 & 0.14$\pm$0.18 &-0.35$\pm$0.10 & 0.06$\pm$0.20 &-0.12$\pm$0.15 &$<$-0.52       &-0.20$\pm$0.30 & -0.08$\pm$0.27 \nl
K      & -2.17$\pm$0.12 &               &-0.01$\pm$0.22 &-0.32$\pm$0.18 & $<$0.10       &-0.02$\pm$0.22 &$<$-0.26       & 0.06$\pm$0.20 &                \nl
O      & -1.91$\pm$0.11 &-0.01$\pm$0.21 &-0.26$\pm$0.17 &-0.33$\pm$0.12 &-0.15$\pm$0.18 & 0.00$\pm$0.17 &$<$-0.61       & 0.50$\pm$0.20 &  0.11$\pm$0.25 \nl
199    & -1.45$\pm$0.11 &-0.17$\pm$0.11 & 0.07$\pm$0.16 &-0.36$\pm$0.08 &-0.27$\pm$0.15 &-0.18$\pm$0.25 &$<$-0.80       &-0.26$\pm$0.25 & -0.34$\pm$0.22 \nl
168    & -2.18$\pm$0.12 &               &-0.64$\pm$0.23 &$<$-0.10       & $<$0.17       & 0.03$\pm$0.22 & $<$0.10       & 0.03$\pm$0.15 &                \nl
\multicolumn{10}{c}{} \nl
\multicolumn{10}{c}{{{\underbar{Sextans}}}} \nl
\multicolumn{10}{c}{} \nl
S35    & -1.93$\pm$0.11 &-0.26$\pm$0.18 &-0.27$\pm$0.20 &-0.46$\pm$0.10 &-0.20$\pm$0.20 & 0.01$\pm$0.20 &$<$-0.74       &-0.15$\pm$0.30 &  0.07$\pm$0.28 \nl
S56    & -1.93$\pm$0.11 &-0.22$\pm$0.15 &-0.18$\pm$0.19 &-0.66$\pm$0.12 &-0.14$\pm$0.20 &-0.08$\pm$0.19 &$<$-0.78       &-0.10$\pm$0.20 &  0.04$\pm$0.28 \nl
S49    & -2.85$\pm$0.13 &               &-0.46$\pm$0.23 &               &               & 0.26$\pm$0.23 &               & $<$0.47       &                \nl
S58    & -1.45$\pm$0.12 &               &-0.29$\pm$0.24 &-0.49$\pm$0.10 &-0.64$\pm$0.30 &-0.33$\pm$0.24 &$<$-0.58       &-0.56$\pm$0.30 & -0.35$\pm$0.38 \nl
S36    & -2.19$\pm$0.12 &               &-0.38$\pm$0.23 &$<$-0.14       & $<$0.10       &-0.24$\pm$0.23 & $<$0.10       &-0.02$\pm$0.25 &                \nl
\enddata
\end{deluxetable}

\clearpage

\begin{deluxetable}{lrrrrrrrrr}
\tablewidth{0pt}
\scriptsize
\tablenum{4 Cont'd}
\tablecaption{Element Abundances for Program Stars}
\tablehead{
\colhead{Star}&
\colhead{[Fe/H]}&
\colhead{[Y/Fe]}&
\colhead{[Ba/Fe]}&
\colhead{[Ce/Fe]}&
\colhead{[Nd/Fe]}&
\colhead{[Sm/Fe]}&
\colhead{[Eu/Fe]}&
\colhead{[Ba/Eu]}&
\colhead{[Ba/Y]} \nl
\colhead{}&
\colhead{(dex)}&
\colhead{(dex)}&
\colhead{(dex)}&
\colhead{(dex)}&
\colhead{(dex)}&
\colhead{(dex)}&
\colhead{(dex)}&
\colhead{(dex)}&
\colhead{(dex)}
}
\startdata
\multicolumn{10}{c}{} \nl
\multicolumn{10}{c}{{{\underbar{M92 = NGC 6341}}}} \nl
\multicolumn{10}{c}{} \nl
III-13 & -2.28$\pm$0.11 &-0.25$\pm$0.07 &-0.50$\pm$0.07 &-0.57$\pm$0.10 &-0.37$\pm$0.10 & 0.08$\pm$0.14 &-0.04$\pm$0.14 &-0.46$\pm$0.16 &-0.25$\pm$0.10 \nl
VII-18 & -2.27$\pm$0.11 &-0.35$\pm$0.07 &-0.50$\pm$0.07 &-0.51$\pm$0.10 &-0.28$\pm$0.10 & 0.02$\pm$0.14 &-0.03$\pm$0.14 &-0.47$\pm$0.16 &-0.15$\pm$0.10 \nl
V-106  & -2.33$\pm$0.11 &-0.33$\pm$0.07 &-0.40$\pm$0.07 &-0.50$\pm$0.10 &-0.14$\pm$0.10 & 0.12$\pm$0.14 & 0.27$\pm$0.14 &-0.67$\pm$0.16 &-0.07$\pm$0.10 \nl
III-65 & -2.30$\pm$0.11 &-0.41$\pm$0.07 &-0.43$\pm$0.07 &-0.47$\pm$0.10 &-0.16$\pm$0.10 & 0.27$\pm$0.14 & 0.16$\pm$0.14 &-0.59$\pm$0.16 &-0.02$\pm$0.10 \nl
\multicolumn{10}{c}{} \nl
\multicolumn{10}{c}{{{\underbar{M3 = NGC 5272}}}} \nl
\multicolumn{10}{c}{} \nl
III-28 & -1.70$\pm$0.11 & 0.08$\pm$0.07 & 0.08$\pm$0.07 &-0.15$\pm$0.10 & 0.28$\pm$0.10 & 0.63$\pm$0.14 & 0.48$\pm$0.14 &-0.40$\pm$0.16 & 0.00$\pm$0.10 \nl
I-21   & -1.44$\pm$0.11 &-0.16$\pm$0.07 & 0.05$\pm$0.07 &-0.19$\pm$0.10 & 0.17$\pm$0.10 & 0.29$\pm$0.14 & 0.38$\pm$0.14 &-0.33$\pm$0.16 & 0.21$\pm$0.10 \nl
IV-101 & -1.44$\pm$0.11 &-0.08$\pm$0.07 & 0.14$\pm$0.07 & 0.06$\pm$0.10 & 0.29$\pm$0.10 & 0.47$\pm$0.14 & 0.40$\pm$0.14 &-0.26$\pm$0.16 & 0.22$\pm$0.10 \nl
\multicolumn{10}{c}{} \nl
\multicolumn{10}{c}{{{\underbar{NGC 2419}}}} \nl
\multicolumn{10}{c}{} \nl
RH10   & -2.32$\pm$0.11 & 0.03$\pm$0.10 &-0.15$\pm$0.10 &-0.05$\pm$0.14 & 0.19$\pm$0.14 & $<$0.80       &               &               &-0.18$\pm$0.14 \nl
\multicolumn{10}{c}{} \nl
\multicolumn{10}{c}{{{\underbar{Draco}}}} \nl
\multicolumn{10}{c}{} \nl
11     & -1.72$\pm$0.11 &-0.68$\pm$0.10 & 0.11$\pm$0.10 &-0.13$\pm$0.14 & 0.23$\pm$0.14 & 0.36$\pm$0.20 & 0.55$\pm$0.20 &-0.44$\pm$0.22 & 0.79$\pm$0.14 \nl
343    & -1.86$\pm$0.11 &-0.45$\pm$0.10 & 0.09$\pm$0.10 &-0.53$\pm$0.14 &-0.19$\pm$0.14 & 0.18$\pm$0.20 & 0.51$\pm$0.20 &-0.42$\pm$0.22 & 0.54$\pm$0.14 \nl
473    & -1.44$\pm$0.07 &-0.74$\pm$0.16 &-0.01$\pm$0.16 & 0.03$\pm$0.35 & 0.50$\pm$0.35 &               &               &               & 0.73$\pm$0.23 \nl
267    & -1.67$\pm$0.13 &-0.73$\pm$0.16 & 0.41$\pm$0.16 & 0.19$\pm$0.35 & 0.11$\pm$0.18 &               &               &               & 1.14$\pm$0.23 \nl
24     & -2.36$\pm$0.09 &$<$-0.65       &-1.19$\pm$0.25 &$<$0.00        &$<$-0.04       &               &               &               &               \nl
119    & -2.97$\pm$0.15 &$<$-0.39       &$<$-1.17       &$<$0.41        & $<$0.37       &               &               &               &               \nl
\multicolumn{10}{c}{} \nl
\multicolumn{10}{c}{{{\underbar{Ursa Minor}}}} \nl
\multicolumn{10}{c}{} \nl
177    & -2.01$\pm$0.11 &-0.52$\pm$0.10 &-0.22$\pm$0.10 &-0.51$\pm$0.14 &-0.09$\pm$0.14 & 0.48$\pm$0.20 & 0.26$\pm$0.20 &-0.48$\pm$0.22 & 0.30$\pm$0.14 \nl
297    & -1.68$\pm$0.11 &-0.04$\pm$0.09 & 0.15$\pm$0.09 &-0.23$\pm$0.11 & 0.45$\pm$0.11 & 1.06$\pm$0.18 & 0.74$\pm$0.18 &-0.59$\pm$0.20 & 0.19$\pm$0.13 \nl
K      & -2.17$\pm$0.12 & 0.06$\pm$0.09 & 1.37$\pm$0.09 & 1.36$\pm$0.11 & 1.28$\pm$0.11 & 1.73$\pm$0.18 & 1.04$\pm$0.18 & 0.33$\pm$0.20 & 1.31$\pm$0.13 \nl
O      & -1.91$\pm$0.11 &-0.67$\pm$0.10 &-0.39$\pm$0.10 &-1.02$\pm$0.14 &-0.27$\pm$0.14 & 0.28$\pm$0.20 & 0.15$\pm$0.20 &-0.54$\pm$0.22 & 0.28$\pm$0.14 \nl
199    & -1.45$\pm$0.11 & 0.34$\pm$0.09 & 0.77$\pm$0.09 & 1.03$\pm$0.11 & 1.13$\pm$0.11 & 1.75$\pm$0.18 & 1.49$\pm$0.18 &-0.72$\pm$0.20 & 0.43$\pm$0.13 \nl
168    & -2.18$\pm$0.12 &-0.34$\pm$0.10 & 0.18$\pm$0.09 &-0.18$\pm$0.14 & 0.23$\pm$0.14 & $<$0.65       & $<$1.30       &               & 0.52$\pm$0.13 \nl
\multicolumn{10}{c}{} \nl
\multicolumn{10}{c}{{{\underbar{Sextans}}}} \nl
\multicolumn{10}{c}{} \nl
S35    & -1.93$\pm$0.11 &-0.07$\pm$0.10 & 0.70$\pm$0.10 & 0.89$\pm$0.11 & 0.79$\pm$0.11 & 1.58$\pm$0.18 &               &               & 0.77$\pm$0.14 \nl
S56    & -1.93$\pm$0.11 &-0.37$\pm$0.13 & 0.23$\pm$0.14 &-0.25$\pm$0.14 & 0.22$\pm$0.14 & 0.86$\pm$0.20 &               &               & 0.60$\pm$0.19 \nl
S49    & -2.85$\pm$0.13 &$<$-0.35       &-1.05$\pm$0.15 & $<$0.45       & $<$0.50       & $<$0.94       &               &               &               \nl
S58    & -1.45$\pm$0.12 &-0.77$\pm$0.13 & 0.11$\pm$0.14 &-0.54$\pm$0.14 & 0.00$\pm$0.14 & 0.52$\pm$0.20 &               &               & 0.88$\pm$0.19 \nl
S36    & -2.19$\pm$0.12 & 0.25$\pm$0.13 & 0.39$\pm$0.14 & $<$0.46       & 0.52$\pm$0.14 & $<$1.34       &               &               & 0.14$\pm$0.19 \nl
\enddata
\end{deluxetable}

\clearpage

\clearpage
\plotone{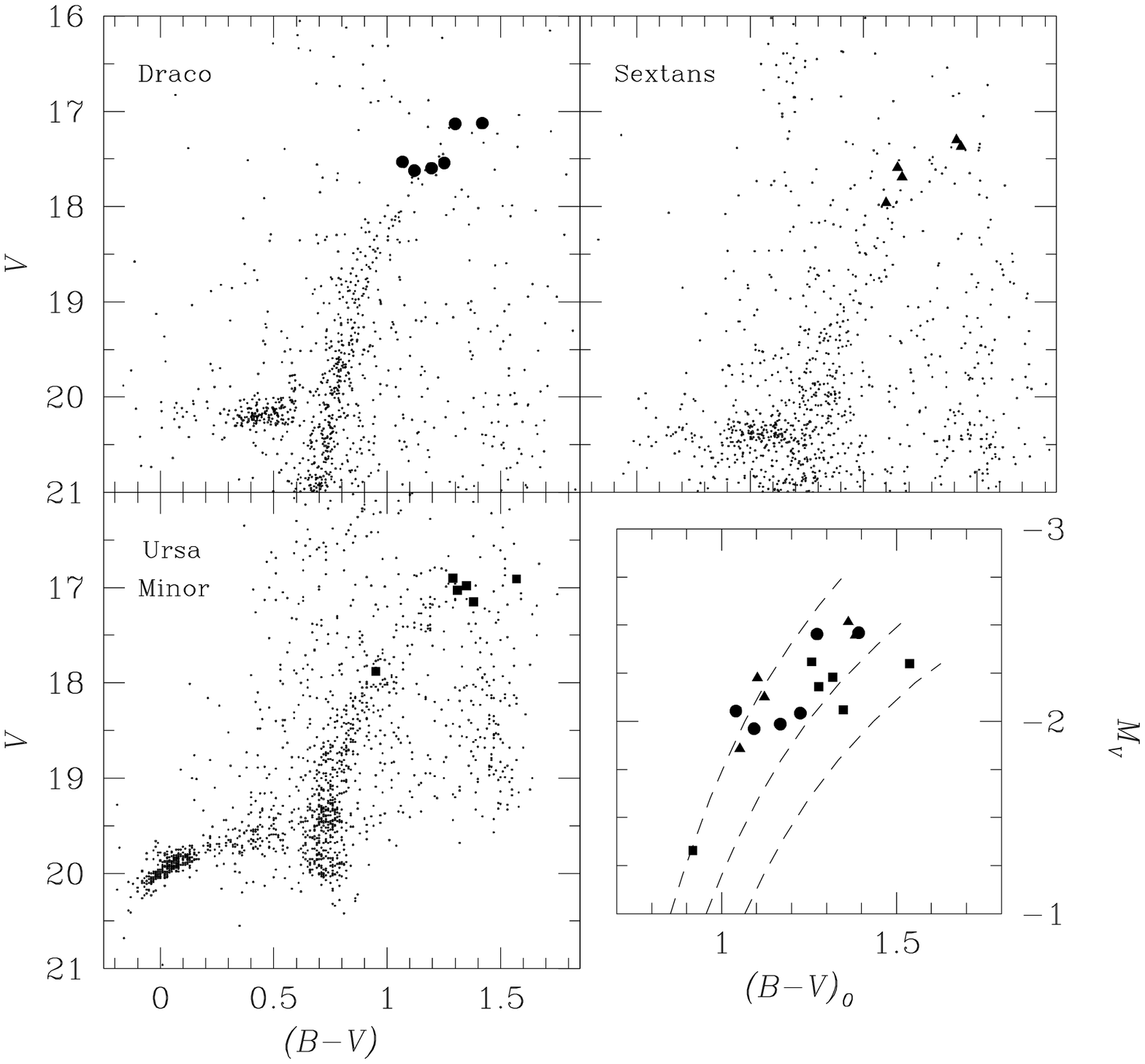}
\figcaption[scs1.ps]{$BV$ color magnitude diagrams for the three dSph galaxies in
our program: Draco, Sextans and Ursa Minor. The red giant branch stars for
which we have HIRES spectra are indicated by the circles (Draco), triangles (Sextans),
and squares (Ursa Minor). The four Draco stars observed  by Shetrone et al. (1998)
are also shown in the first panel. Photometric data for Draco,
Sextans and Ursa Minor are from Stetson (2000), Suntzeff et al. (1993) 
and Cudworth et al. (2000), respectively. The lower right panel shows the 
location of our program stars in the $M_V$-$(B-V)_0$ plane. Dashed lines show 
13 Gyr isochrones from Bergbusch \& VandenBerg (1992) having metallicities of 
[Fe/H] = $-2.26$, $-1.66$ and $-1.26$ dex.  The symbol are the same as in the 
proceeding panels.
\label{fig1}}
\clearpage

\clearpage
\plotone{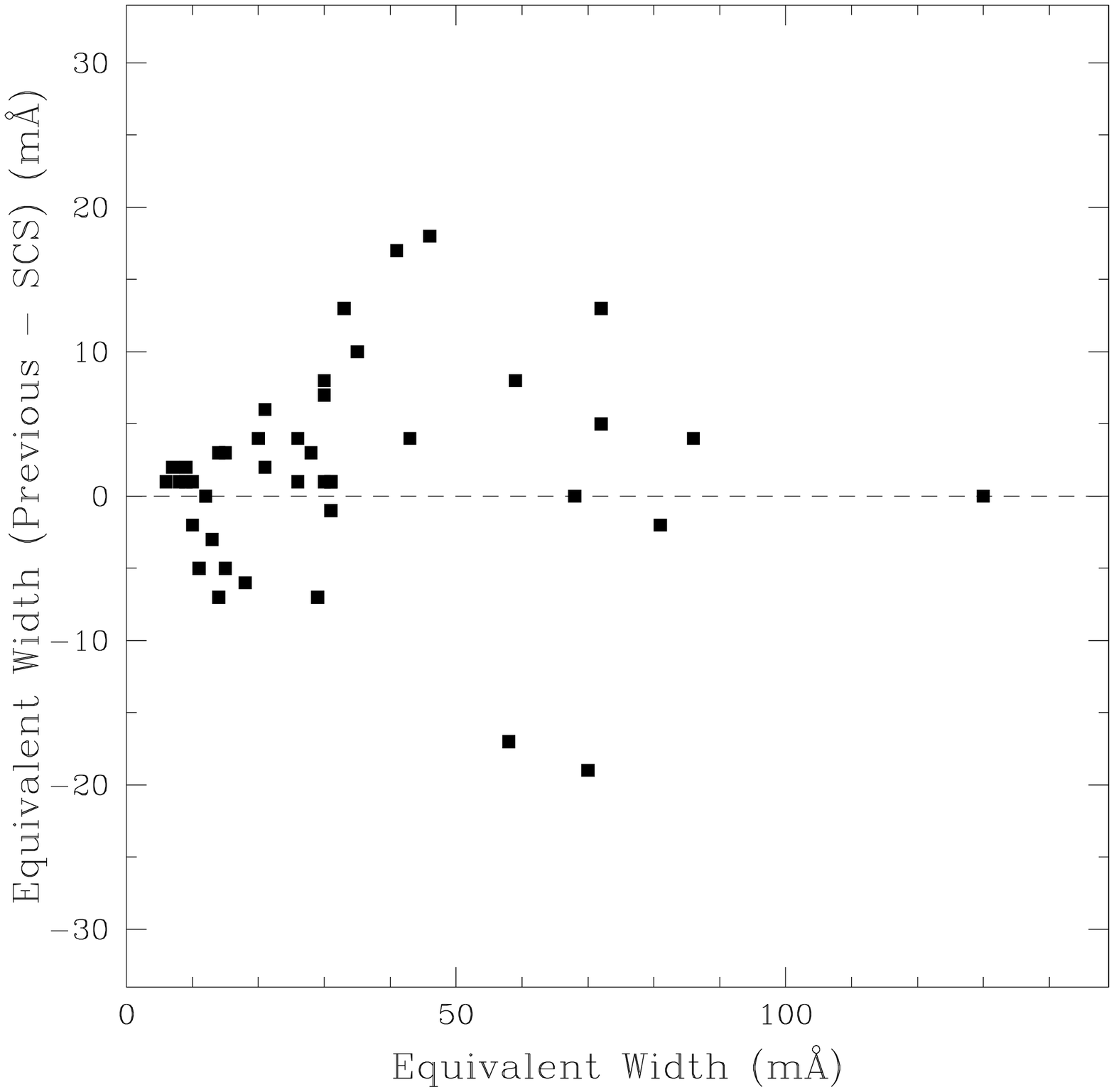}
\figcaption[scs2.ps]{Comparison of equivalent widths for the globular
cluster red giants presented in Table 3 with those previously reported by 
Sneden et al. (1991) and Kraft et al. (1992).
\label{fig2}}
\clearpage

\clearpage
\plotone{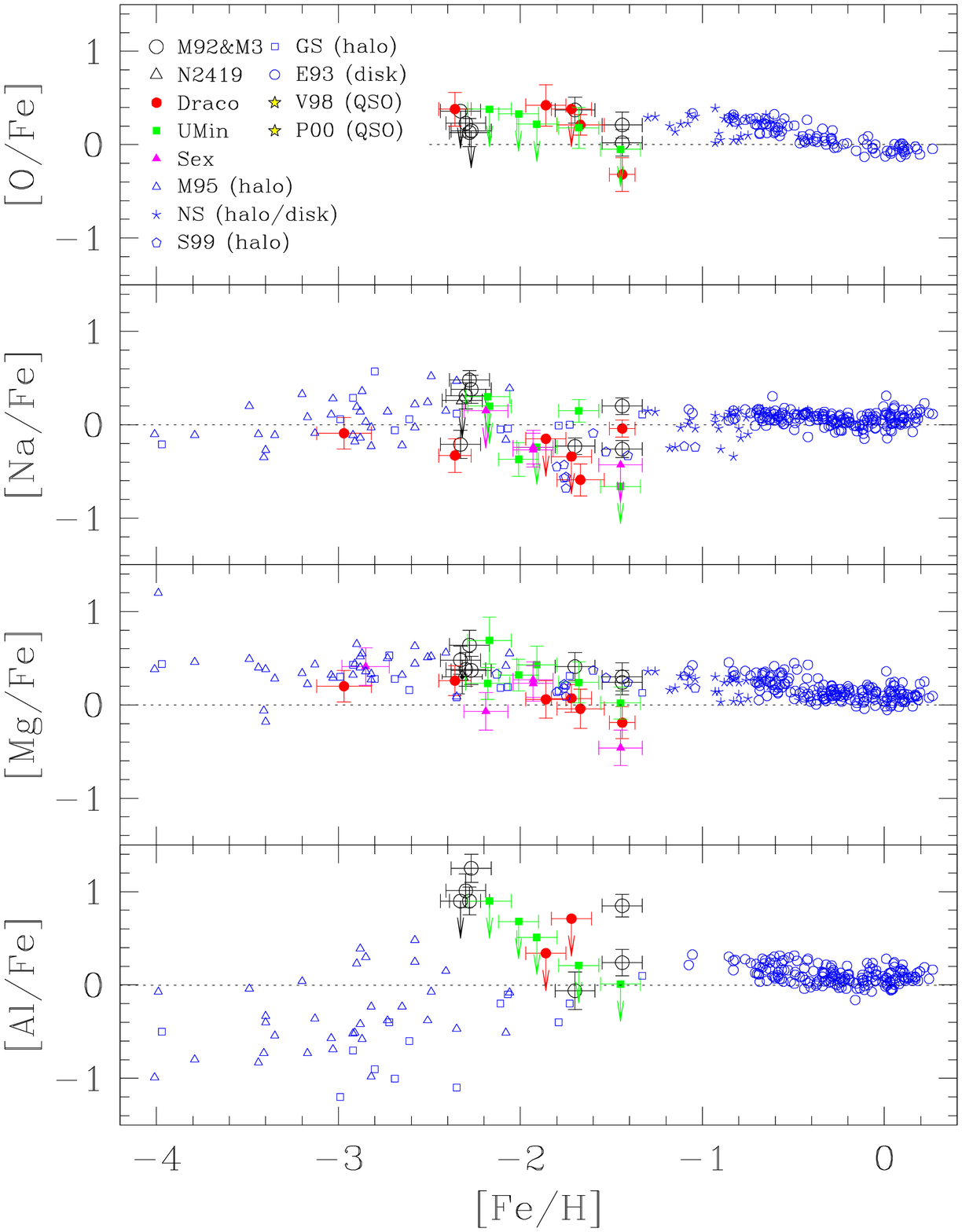}
\figcaption[scs3.ps]{Abundance ratios for light (Na, Al) and alpha 
(O, Mg) elements, plotted against logarithmic iron-to-hydrogen ratio, [Fe/H]. 
The dashed line shows the solar ratios. The key to the symbols is given in the 
upper panel. The halo field star samples are from McWilliam et al. (1995; M95),
Nissen \& Schuster (1997; NS), Stephens (1999; S99), and the series of
papers by Gratton (1989), Sneden et al. (1991) and 
Gratton \& Sneden (1988; 1991; 1994; GS). 
Disk stars are from Edvardsson et al. (1993; E93) and Nissen \& Schuster (1997; NS).
Element abundances for damped Ly$\alpha$ absorbers in the direction
of several QSOs are taken from Vladilo (1998; V98) and
Pettini et al. (2000; P00).
\label{fig3}}
\clearpage

\clearpage
\plotone{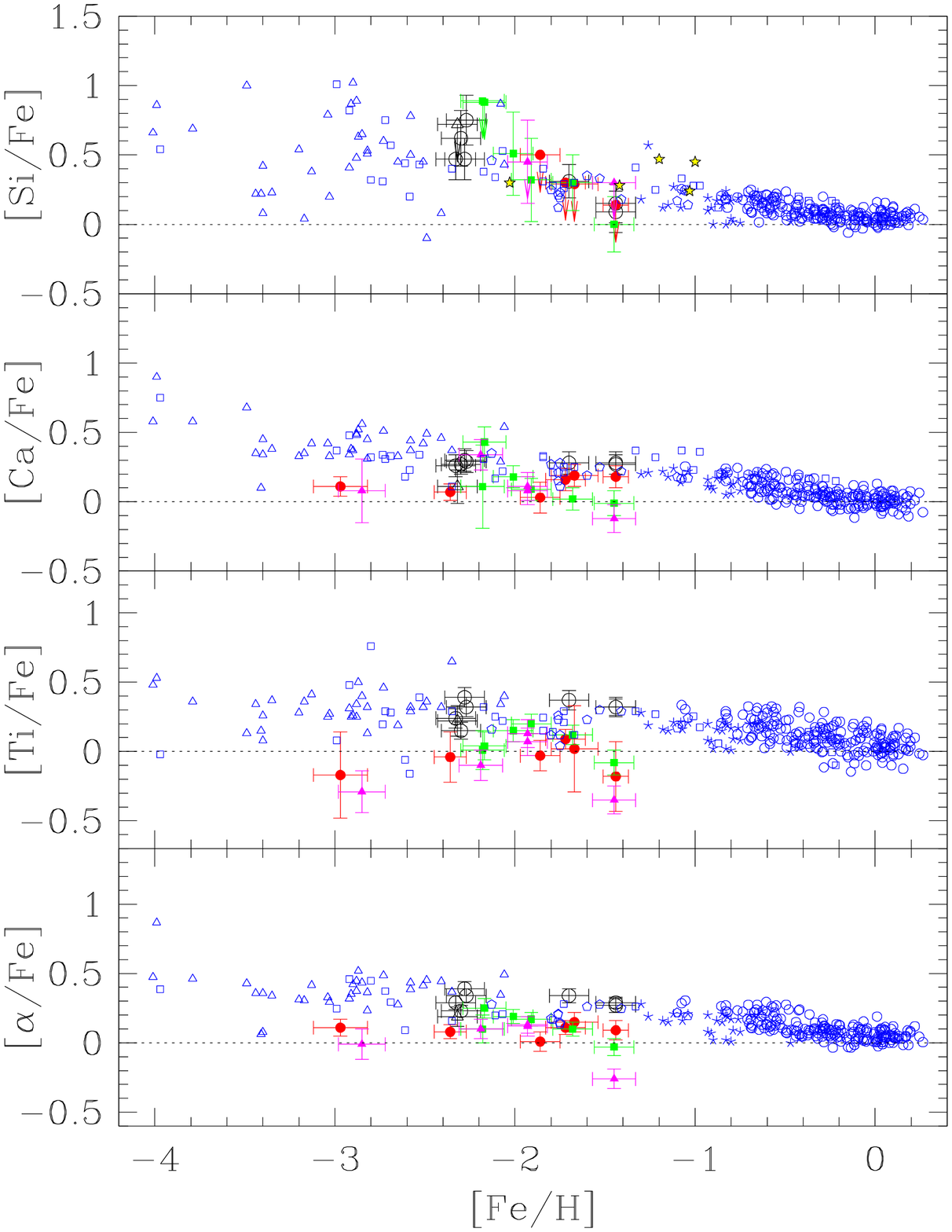}
\figcaption[scs4.ps]{Abundance ratios for additional alpha elements (Si, Ca, Ti) 
plotted against [Fe/H]. The lower
panel shows the {\it mean} alpha element abundance ratio, taken here as 
[${\alpha}$/Fe] = ${1 \over 3}$([Mg/Fe + Ca/Fe + Ti/Fe]).
The symbols are the same as in Figure 3.
\label{fig4}}
\clearpage

\clearpage
\plotone{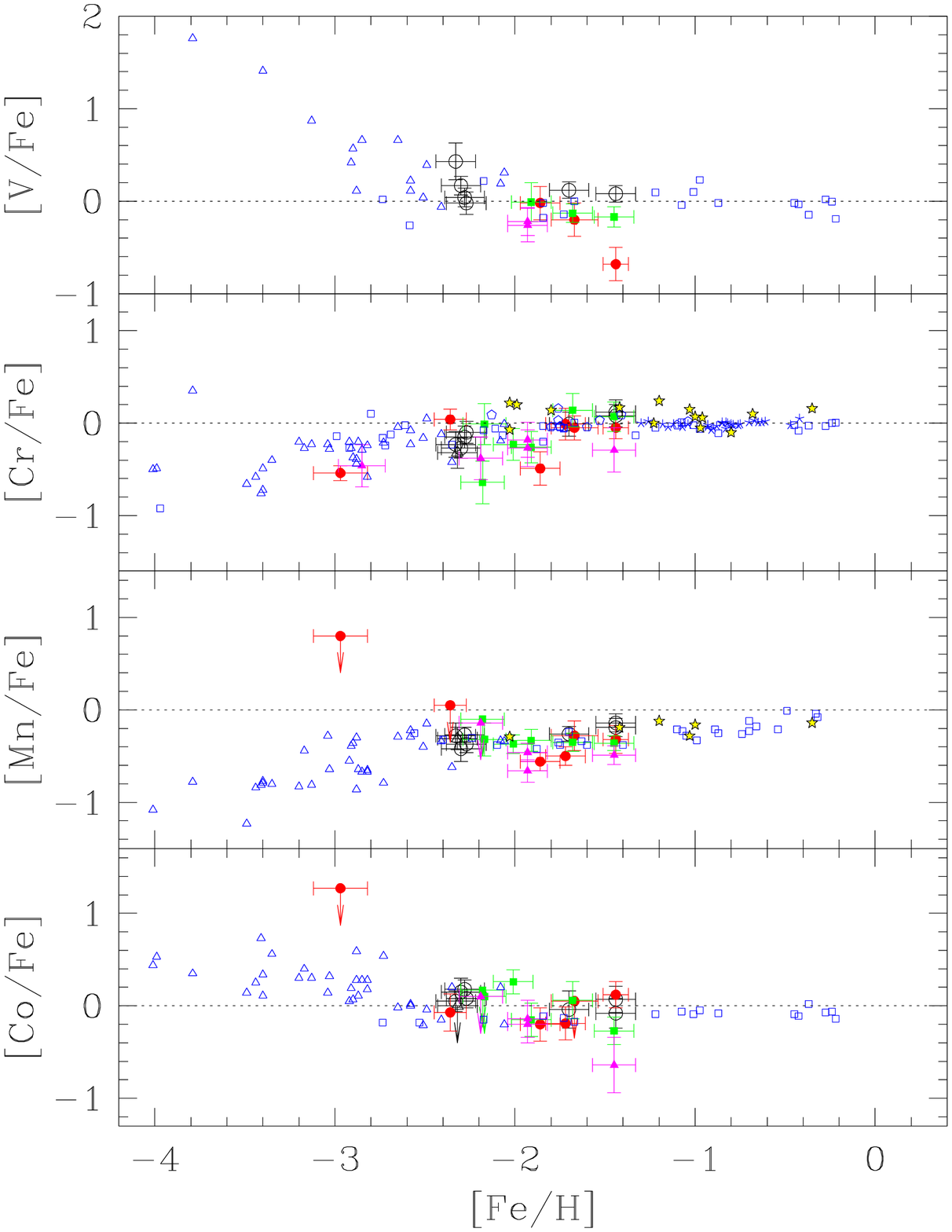}
\figcaption[scs5.ps]{Abundance ratios for iron peak elements (V, Cr, Mn, Co)
plotted against [Fe/H]. The symbols are the same 
as in Figure 3.
\label{fig5}}
\clearpage

\clearpage
\plotone{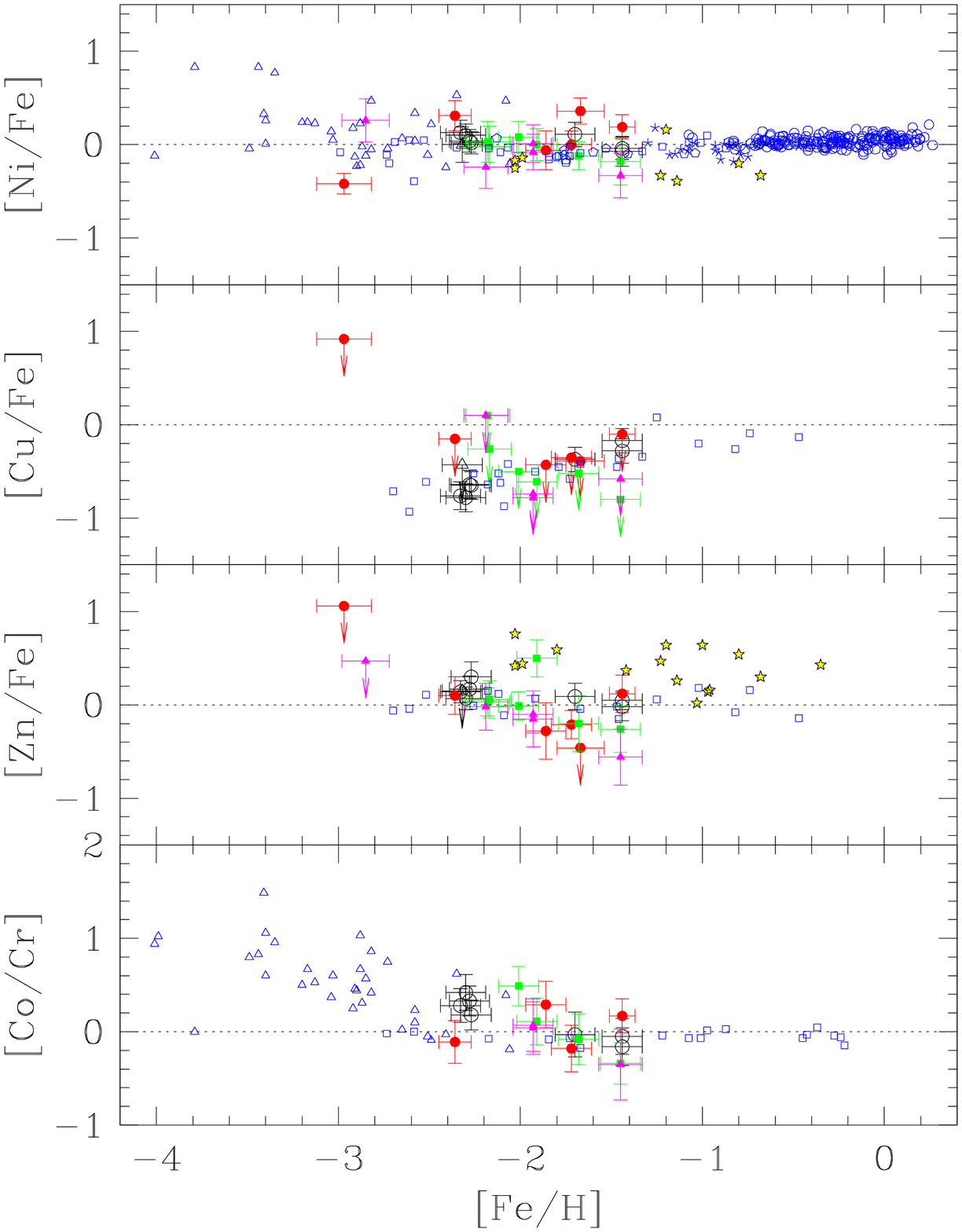}
\figcaption[scs6.ps]{Abundance ratios for additional iron peak elements (Ni, Cu, Zn)
plotted against [Fe/H]. The lower panel shows the
dependence of [Co/Cr] on metallicity. The symbols are the same as in Figure 3.
\label{fig6}}
\clearpage

\clearpage
\plotone{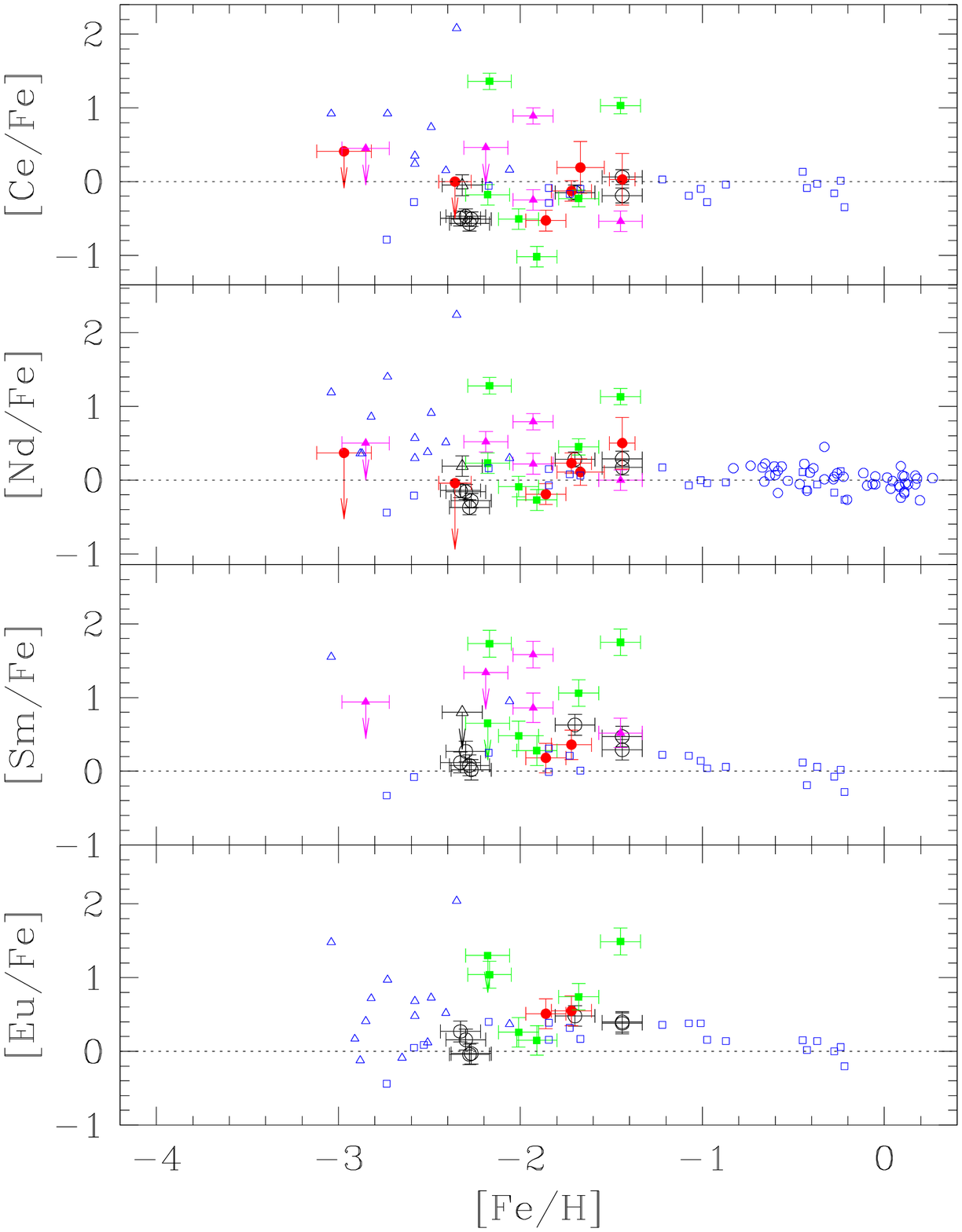}
\figcaption[scs7.ps]{Abundance ratios for heavy elements (Ce, Nd, Sm, Eu) plotted
against [Fe/H]. The symbols are the same as in Figure 3.
\label{fig7}}
\clearpage

\clearpage
\plotone{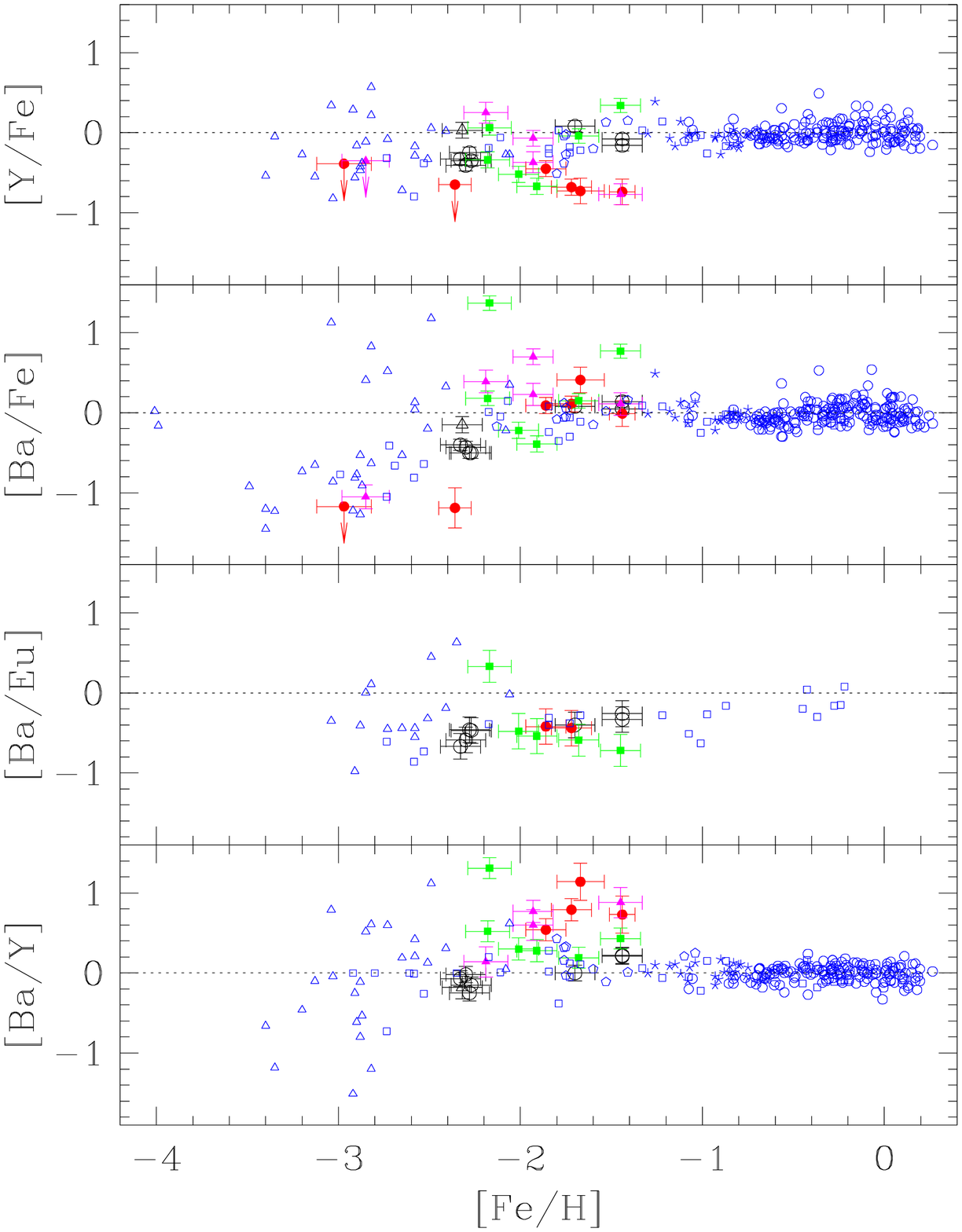}
\figcaption[scs8.ps]{Abundance ratios for neutron-capture elements (Y, Ba) are
plotted against [Fe/H] in the upper two panels.
The lower two panels show the dependence of [Ba/Eu] and [Ba/Y] on metallicity.
The symbols are the same as in Figure 3.
\label{fig8}}
\clearpage

%
 
\end{document}